\documentclass[a4paper,11pt]{article}
\pdfoutput=1 % if you are submitting a pdflatex (i.e. if you have images in pdf, png, or jpg format)
\usepackage{jcappub} % for details on the use of the package, please see the JCAP-author-manual
\usepackage[T1]{fontenc} % if needed
\usepackage{units} % quantities with units
\usepackage[capitalize]{cleveref} % clever references
\crefformat{figure}{figure #2#1#3}
\crefformat{section}{section #2#1#3}
\crefformat{equation}{equation #2#1#3}
\usepackage[dvipsnames]{xcolor} % more colors
\usepackage{float} % to force pictures placement
\usepackage{booktabs} % options for tables
\usepackage{todonotes} % todo notes
\usepackage[switch*,modulo]{lineno} % line numbers
\usepackage{amsmath,amssymb} % extra math symbols
\usepackage{xspace} % to draw the offline logo
\usepackage{subfigure} % to use subfigure command
\usepackage{wrapfig} % to use wrapfigure command
\usepackage{makecell} % to handle long text in a tabular cell
\usepackage{subcaption} % to handle subcaption
\newcommand{\Offline}{\mbox{$\overline{\rm Off}$\hspace{.05em}\raisebox{.3ex}{$\underline{\rm line}$}}\xspace}

\title{\boldmath Search for a diffuse flux of photons with energies above tens of PeV at the Pierre Auger Observatory}
\author[13]{A.~Abdul Halim,}
\author[70]{P.~Abreu,}
\author[53,51]{M.~Aglietta,}
\author[1]{I.~Allekotte,}
\author[78,77]{K.~Almeida Cheminant,}
\author[7,12]{A.~Almela,}
\author[44,45]{R.~Aloisio,}
\author[76]{J.~Alvarez-Mu\~niz,}
\author[44]{A.~Ambrosone,}
\author[76]{J.~Ammerman Yebra,}
% ...

\affiliation[1]{Centro At\'omico Bariloche and Instituto Balseiro (CNEA-UNCuyo-CONICET), San Carlos de Bariloche, Argentina}
% ...
\affiliation[7]{Instituto de Tecnolog\'\i{}as en Detecci\'on y Astropart\'\i{}culas (CNEA, CONICET, UNSAM), Buenos Aires, Argentina}
% ...
\affiliation[12]{Universidad Tecnol\'ogica Nacional -- Facultad Regional Buenos Aires, Buenos Aires, Argentina}
\affiliation[13]{University of Adelaide, Adelaide, S.A., Australia}
% ...
\affiliation[44]{Gran Sasso Science Institute, L'Aquila, Italy}
\affiliation[45]{INFN Laboratori Nazionali del Gran Sasso, Assergi (L'Aquila), Italy}
% ...
\affiliation[51]{INFN, Sezione di Torino, Torino, Italy}
% ...
\affiliation[53]{Osservatorio Astrofisico di Torino (INAF), Torino, Italy}
% ...
\affiliation[70]{Laborat\'orio de Instrumenta\c{c}\~ao e F\'\i{}sica Experimental de Part\'\i{}culas -- LIP and Instituto Superior T\'ecnico -- IST, Universidade de Lisboa -- UL, Lisboa, Portugal}
% ...
\affiliation[76]{Instituto Galego de F\'\i{}sica de Altas Enerx\'\i{}as (IGFAE), Universidade de Santiago de Compostela, Santiago de Compostela, Spain}
\affiliation[77]{IMAPP, Radboud University Nijmegen, Nijmegen, The Netherlands}
\affiliation[78]{Nationaal Instituut voor Kernfysica en Hoge Energie Fysica (NIKHEF), Science Park, Amsterdam, The Netherlands}
% ...
\affiliation[c]{School of Physics and Astronomy, University of Leeds, Leeds, United Kingdom}
% ...

\emailAdd{spokespersons@auger.org}

\abstract{Diffuse photons of energy above \unit[0.1]{PeV}, produced through the interactions between cosmic rays and either interstellar matter or background radiation fields, are powerful tracers of the distribution of cosmic rays in the Galaxy.  Furthermore, the measurement of a diffuse photon flux would be an important probe to test models of super-heavy dark matter decaying into gamma-rays. In this work, we search for a diffuse photon flux in the energy range between \unit[50]{PeV} and \unit[200]{PeV} using data from the Pierre Auger Observatory. For the first time, we combine the air-shower measurements from a \unit[2]{km$^{2}$} surface array consisting of $19$ water-Cherenkov surface detectors, spaced at \unit[433]{m}, with the muon measurements from an array of buried scintillators placed in the same area. Using \unit[15]{months} of data, collected while the array was still under construction, we derive upper limits to the integral photon flux ranging from $13.3$ to \unit[$13.8$]{km$^{-2}$\,sr$^{-1}$\,yr$^{-1}$} above tens of PeV. We extend the Pierre Auger Observatory photon search program towards lower energies, covering more than three decades of cosmic-ray energy. This work lays the foundation for future diffuse photon searches: with the data from the next 10 years of operation of the Observatory, this limit is expected to improve by a factor of $\sim20$.}

\keywords{cosmic rays, ultra-high-energy photons, Pierre Auger Observatory, surface detector, underground muon detector}
\arxivnumber{2502.02381}

\begin{document}
\maketitle
\flushbottom

\section{Introduction}
\label{sec:intro}

%Context - We explain why this research is important

%Astrophysics, cosmic-ray sources (PeVatrons)
The origin and acceleration of very-high-energy (VHE, $E \gtrsim \unit[10^{14}]{eV}$) cosmic rays can be investigated through the detection of photons produced by the interaction between cosmic rays and surrounding matter near their sources~\citep{Bhatta2022,TibetAS2019,Gopal2010}. Photons with energies between \unit[$10^{14}$]{eV} and \unit[$10^{18}$]{eV} interact with background radiation fields, limiting their travel to at most a few megaparsecs (Mpc)~\citep{Settimo2015}, making them ideal probes for studying sources within our Galaxy and its vicinity. Recent observations have identified primary photons from Galactic sources with energies reaching up to \unit[$\sim10^{15}$]{eV}~\citep{LHAASO2021,HAWC2021}.

%\subsection{Theoretical predictions of diffuse photon flux}
%Diffuse cosmogenic flux from CR propagation
In addition to astrophysical sources, cosmogenic photons are expected to arise from interactions between cosmic rays and the cosmic microwave background, as well as the extragalactic background light, in intergalactic space~\citep{Gelmini2022,Bobrikova2021,Batista2019}.
%Specifically, when UHE protons interact with background photons, they produce neutral and charged pions through the resonant production of a $\Delta^{+}$, which then decay into secondary photons, electrons, and neutrinos~\citep{Greisen1966,G.Zatsepin1966}. The flux of secondary photons, which carry about $10\%$ of the energy of the primary particle, is estimated to be between five and seven orders of magnitude lower than the measured cosmic-ray flux for energies above \unit[$10^{17}$]{eV}. This diffuse component is influenced by various factors, such as the energy spectrum and composition of the UHE cosmic rays at their sources, their propagation distance, and the intervening background fields~\citep{Decerprit2011,G.Gelmini2008}. 
Light particles have higher interaction cross-sections with these fields, leading to more frequent production of secondary photons than occurs with heavier nuclei~\citep{Hooper2011_2}. Recent studies using the data acquired by the Large High Altitude Air Shower Observatory (LHAASO) suggest that the average cosmic-ray mass is heavier than helium at around \unit[$10^{16}$]{eV} with an increasing trend towards heavier elements up to about \unit[$2\times10^{17}$]{eV}~\citep{Cao2024}, after which the composition either remains constant or becomes lighter~\citep{Auger2024b,Auger2024c,Yushkov2019,Fujita2023,Plum2022,Knurenko2019}.

%The mass composition of cosmic rays around \unit[$10^{17}$]{eV} has been extensively studied using various experimental facilities. The findings generally indicate a mixed composition around \unit[$10^{17}$]{eV}. Recent studies using the data acquired by the Large High Altitude Air Shower Observatory (LHAASO) suggest that the average mass is heavier than helium at around \unit[$10^{16}$]{eV} with an increasing trend towards heavier elements up to about \unit[$2\times10^{17}$]{eV}~\citep{Cao2024}, after which the composition either remains constant or becomes lighter with increasing energies~\citep{Fujita2023,Plum2022,Knurenko2019}. The absolute scale of the mass composition depends on the choice of the hadronic interaction model used in air-shower simulations~\citep{Aartsen2019}.

%The elemental fractions align well with phenomenological models of the transition region between galactic and extragalactic cosmic rays, where heavier elements retain a harder spectral index to higher energies~\citep{Gaisser2012}.

%Based on the study of electromagnetic and muonic air shower components, data acquired by KASCADE-Grande exhibits a knee-like feature in the Si+Fe component of the cosmic-ray spectrum accompanied by a flattening of the H+He+CNO component around \unit[$10^{16.7}$]{eV}~\citep{KASCADE2023}. The former feature has been correlated to the observed break in the all-particle cosmic-ray spectrum known as the second knee~\citep{TA2018}.  

%Diffuse flux from CR interaction with Galactic matter
Another significant contribution to the diffuse photon flux comes from the interactions between VHE cosmic rays and Galactic disk matter~\citep{Lipari2018,Berat2022}. Similar to fluxes from cosmic-ray propagation through radiation fields, this component depends on the flux and composition of the primary cosmic rays, as well as the distribution of the gas in the Galactic disk, and the interaction cross-sections. This flux diminishes as $E^{-2}$ and may become the dominant component of the total cosmogenic photon flux below \unit[$10^{17}$]{eV}, though it remains four to five orders of magnitude lower than the energy-integrated cosmic-ray flux. Measurements from LHAASO provide further insight into the diffuse gamma-ray emissions from the Galactic plane, spanning energies between \unit[$10^{13}$]{eV} and \unit[$10^{15}$]{eV}. These measurements indicate that the gamma-ray flux is three times higher than predictions based on local cosmic-ray interactions with Galactic matter, particularly in the inner Galactic plane~\citep{LHAASO2023}. These results suggest additional emission sources or spatial variations in cosmic-ray fluxes, pointing to a more complex picture of gamma-ray production in the Milky Way. Upper limits on the diffuse photon flux between \unit[$10^{15}$]{eV} and \unit[$1.5\times10^{17}$]{eV} have been set from the Northern Hemisphere using experimental facilities, including KASCADE-Grande~\citep{KASCADEGrande2017}, EAS-MSU~\citep{Formin2017} and CASA-MIA~\citep{CASAMIA1997}, while a diffuse search towards the Galactic plane has been performed with data measured by IceTop at \unit[$2\times10^{15}$]{eV}~\citep{IceCubePeVPh2020}. 

%Diffuse flux from pp interactions in the halo 
%Although the density of gas in the Galactic halo is much lower than in the disk, it is sufficient to contribute to the diffuse photon flux through proton-proton interactions. Assuming that the observed IceCube astrophysical neutrino flux originates entirely from the charged pions produced in such interactions, the associated photon flux has been estimated for various source distributions and spectral shape of the neutrino flux~\citep{Kalashev2014}. However, this component is significantly constrained by the current observational photon limits above \unit[$10^{17}$]{eV}~\citep{Auger2022c}.

%Diffuse flux from dark matter decay
The diffuse photon flux can also be used to constrain the lifetime of super-heavy dark matter (SHDM) within the Galactic center. The decay of SDHM may contribute to the diffuse photon flux above a few \unit[$10^{15}$]{eV}~\citep{Auger2023a,Anchordoqui2021,Kachelriess2018}. Since photons generated in the Galactic center are not significantly attenuated owing to the source proximity, they provide a prime signal for probing gamma-ray production across various SHDM decay channels~\citep{Aloisio2015,Chianese2021}.

%The Galactic center is of special interest also due to the gamma-ray flux measured by the H.E.S.S. Collaboration up to about \unit[5$\times10^{13}$]{eV} with no observed cutoff or a spectral break~\citep{HESS2016}. Thus, detecting photon signatures at higher energies may help determine the feasibility of the supermassive black hole in the Galactic center as a source of cosmic rays~\citep{Aharonian2013}, which has been proposed as a potential accelerator up to $\sim$\unit[$10^{15}$]{eV} assuming a higher accretion rate in recent epochs~\citep{Aharonian2005}.

%The lifetime-and-mass parameter space of such exotic particles can be constrained by accounting for the current limits to the integral UHE photon flux~\citep{}

%Differences between photonic and hadronic air showers
The searches for photons at ultra-high energies (UHE, $E \gtrsim \unit[10^{17}]{eV}$) are performed by measuring extensive air showers, i.e., cascades of secondary particles produced in the atmosphere. The main challenge in these searches is distinguishing primary photons from the overwhelming background of charged cosmic rays. The separation is based on air-shower properties: showers initiated by photon primaries develop almost entirely through electromagnetic processes, while those initiated by hadrons contain a much larger number of muons~\citep{Grieder2010,Bluemer2009,Risse2007}. Due to the smaller multiplicity of electromagnetic compared to hadronic interactions, the atmospheric depth of maximum shower development, $X_\text{max}$, is expected to be deeper for a photon than for a hadronic primary. Also, taking into account the muonic component, this leads to a flatter lateral spread of the shower for hadronic cosmic rays compared with primary photons. These distinctive characteristics of photon showers are used to discriminate them from hadronic showers.

%\subsection{Search for primary photons from the Southern hemisphere}. Auger previous photon searches. Gaps in knowledge: search for photons from the Southern hemisphere
The Pierre Auger Observatory~\citep{Collaboration2015f} integrates multiple detection techniques to extract information from air showers reaching the ground using a surface detector (SD) and from the measurement of the light emitted by air showers using a fluorescence detector (FD). The Pierre Auger Collaboration, combining SD and FD measurements, has set the most stringent upper limits on the integral photon flux at energies above \unit[$2\times10^{17}$]{eV}~\citep{Auger2022c,Auger2024g,Auger2023c}. These analyses rely either on the rise-time and the integrated signal measured by water-Cherenkov detectors (WCDs) within the SD stations~\citep{Collaboration2015f}, the slope of the lateral signal fall-off, the indirect muon number estimation, or the measurement of the fluorescence radiation produced during the air-shower development.

In this study, we extend the search for primary photons done at the Auger Observatory towards lower energies, reaching down to \unit[$5\times10^{16}$]{eV}. The extension is made possible by a dense array of $19$ WCDs deployed over \unit[2]{km$^{2}$} and the direct measurements of the air-shower muonic component by an array of buried scintillators deployed in the same region, as discussed in \cref{sec:detectorsAndData}. The Monte Carlo-driven reconstruction of photon-initiated events and the definition of a unique energy scale to treat both photon and hadronic events are presented in \cref{sec:MCRec}. The photon-hadron discrimination is performed using an observable, $M_b$, defined as the event-wise weighted sum of the muon density measurements. The power of this photon-hadron discriminator is presented in-depth in \cref{sec:discrimination}. The selection criteria considered in data are described in \cref{sec:application}. The results of the photon search are presented in \cref{sec:results}.

%Given that air showers generated by photons are predominantly electromagnetic, muon quantification is a powerful observable for distinguishing primary photons from the hadronic cosmic-ray background. 

%This study will allow us to constrain further the theoretical dark-matter models, the properties of cosmic-ray sources, and the high-energy regime of the Galactic gamma-ray sources, leveraging the privileged exposure of Auger towards the Galactic plane.

\section{Detection systems and data}
\label{sec:detectorsAndData}

%Robust and precise detection systems are essential to tackle the challenging endeavor of detecting UHE photons around \unit[$10^{17}$]{eV}. At such energies, a nested array of surface detector stations separated by \unit[433]{m} is employed to detect UHE cosmic rays at the  Pierre Auger Observatory.

%\subsection[The $433$-m array]{The \boldmath{$433$}-m array}
%\label{sec:433mArrays}

%SD-433: description, station signals, saturated stations
The surface detector of the Pierre Auger Observatory consists of three regular detector grids, spaced by \unit[1500]{m}, \unit[750]{m} and \unit[433]{m}, each designed to probe different regions of the cosmic-ray energy spectrum. Among these, the SD-433~\citep{Silli2021_2,BrichettoICRC2023} is used in this study and comprises of $19$ WCDs arranged in seven regular hexagons, covering an area of \unit[$\sim$2]{km$^{2}$}. Each WCD operates independently, calibrated using atmospheric particles, with signals measured in units of vertical equivalent muons (VEM)~\citep{Bertou2006}. A more detailed description can be found in~\citep{Collaboration2015f,Allekotte2008}.

The SD triggering system starts with low-level triggers from individual WCDs, progressing to the high-level trigger (T5). This trigger is used to select events where the station with the highest signal is surrounded by five (5T5) or six (6T5) active stations present in the ring of nearest neighbors. It ensures an accurate reconstruction of the impact point on the ground and reduces the calculation of the exposure to purely geometrical arguments~\citep{Collaboration2010}. 

%The first two levels enforce threshold trigger conditions, requiring coincidence between two or three PMTs within a specific number of ADC time bins. The T3 trigger identifies time coincidences between the signals in diﬀerent WCDs that might indicate a real air shower. 

\begin{figure}[!tb]
\centering
\includegraphics[width=0.32\textwidth,trim=0 -0.5cm 0 0]{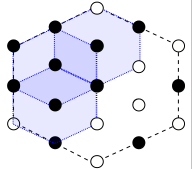} \includegraphics[width=0.49\textwidth]{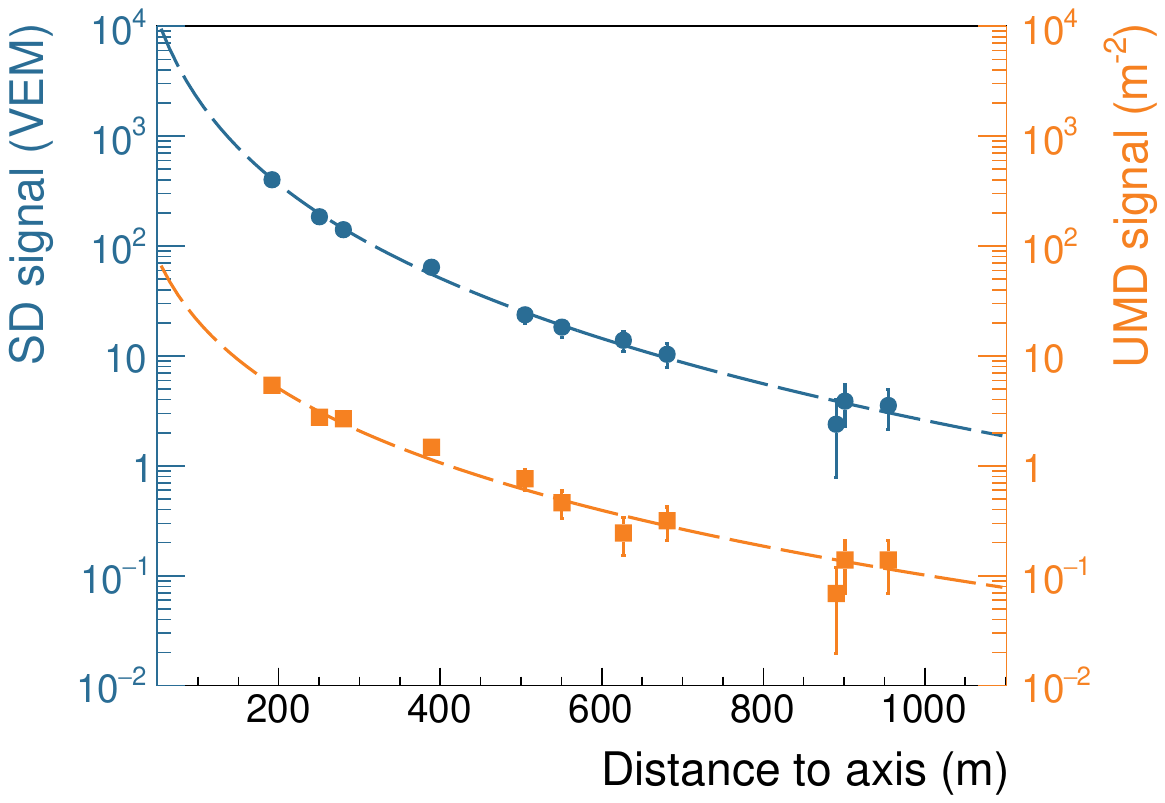}
\caption{Left: A schematic view of the SD-433 array. Solid black positions are equipped with buried scintillators. Events acquired by the three highlighted hexagons are considered for the photon search. Right: One of the events acquired on February 27, 2021, part of the selected data set. The reconstructed energy is \unit[$(4.1\pm0.1)\times10^{17}$]{eV}, obtained with a dedicated data energy scale~\citep{BrichettoICRC2023}, with a zenith angle of \unit[$(26.3\pm0.3)$]{$^\circ$}. Dashed lines indicate the fitted lateral distribution functions (see \cref{sec:MCRec}).}
\label{fig:UMD433}
\end{figure}

%UMD: description, electronics, slave detector
In the configuration used in this work, eleven stations of the SD-433 array are co-located with stations of the Underground Muon Detector (UMD), as illustrated in \cref{fig:UMD433}, left. Each UMD station consists of three modules, each containing $64$ plastic scintillator bars measuring \unit[400]{cm} in length, \unit[4]{cm} in width and \unit[1]{cm} in thickness providing a sensitive area of \unit[$\sim$10]{m$^2$}~\citep{AMIGAFAL2019,Collaboration2016}. The central station on the array's western edge contains six modules, providing a sensitive area of \unit[$\sim$50]{m$^2$}, compared to the standard \unit[30]{m$^2$} area in all other stations. All modules are buried under \unit[2.3]{m} of soil equivalent to a vertical shielding of \unit[540]{g/cm$^2$}, that filters out most of the particles produced in the air except muons with kinetic energies \unit[$\gtrsim$1]{GeV}. When a charged particle passes through the module, the scintillation material emits photons that are collected by a \unit[1.2]{mm} wavelength-shifter optical fiber and conducted to an array of $64$ silicon photomultipliers (SiPMs) Hamamatsu S13081-050CS~\citep{Collaboration2017}. Front-end electronics convert the analog pulses of each SiPM into binary signals, effectively translating the current pulses into a sequence of digital boolean samples. These output signals are sampled by a Field-Programmable Gate Array at \unit[320]{MHz}, corresponding to a sampling interval of \unit[3.125]{ns}, resulting in a binary trace of $2048$ bits stored in the front-end memory. The back-end electronics handles all calibration, control, and monitoring tasks~\citep{Collaboration2021e}. Additionally, the surface electronics, common to all UMD station modules, interfaces with the SD electronics to check for a trigger and retrieve and transfer the binary traces upon an event data request.

%, glued to a lengthwise groove on each bar. The light is then
% based on a comparison with a remotely adjustable voltage threshold
%These samples are stored in the front-end memory, which consists of two circular buffers capable of holding up to \unit[6.4]{$\mu$s} of data. 

%\begin{wrapfigure}[20]{I}{0.4\textwidth}
%\begin{figure}[!tb]
%\centering
%\includegraphics[width=0.4\textwidth]{./images/UMD433}
%\caption{A schematic view of the SD-433 array. Solid black positions are equipped with buried muon detectors. Events acquired by the three highlighted hexagons are considered for the photon search.\label{fig:UMD433}}
%\end{wrapfigure}
%\end{figure}

%UMD: number of muons estimation, muon density
The highly segmented UMD modules are designed to count individual muons when triggered by the associated SD station. The counting procedure, performed on the binary traces, has been optimized to provide an unbiased estimate of the number of muons reaching a module. Specifically, muons are identified along the traces as a compact pattern of four consecutive ``1'' samples. Upon a successful match, an inhibition time window of $12$ samples is activated, preventing additional muon counts within this period~\citep{DeJesusICRC2023,Botti2021a}. The estimated muon density is defined as the number of muons over the projected area of a module to account for the decrease of its sensitive area with the air-shower zenith angle.

%The estimated number of muons is subsequently corrected for the corner-clipping effect, which could lead to systematic overcounting. Two phenomena contribute to this effect: first, muons crossing two adjacent scintillator bars may generate muon patterns in both of them; second, knock-on electrons produced by ionization along the muon track through the soil may generate positive ``1'' samples if produced near the module top surface. 

%\subsection{Event reconstruction}
%\label{sec:eventRec}
%SD-433: LDF
To reconstruct the incoming direction of the primary particle and the lateral distribution function (LDF)~\citep{Auger2020} of the air showers we use the signals from the WCDs. The LDF is described by the Nishimura-Kamata-Greisen (NKG) function~\citep{Kamata1958,Greisen1956} from which we extract the shower size, $S(r_\text{ref})$, defined as the value at an optimal distance, $r_\text{ref}$. For an array spacing of \unit[433]{m}, $r_\text{ref}$ is chosen as \unit[250]{m} to minimize the uncertainties of the signal due to the imperfect knowledge of the functional form of the LDF in individual events~\citep{Newton2007}. The shower size, $S(250)$, corrected for the attenuation in the atmosphere serves as an observable that can be calibrated against the primary energy. Although for most events detected with SD-433 (hadronic showers) a data-driven energy calibration is employed~\citep{BrichettoICRC2023,Dembinski2016}, the energy calibration for photon events is instead performed using detailed Monte Carlo simulations, as discussed in \cref{sec:MCRec}.

%Since shower development varies with the mass composition of the primary cosmic ray (e.g.,\, photonic or hadronic particle), the energy calibration is performed for each scenario using detailed simulations of the showers and the detectors, as discussed in \cref{sec:MCRec}.

%\subsection{Data selection}
%\label{sec:dataSelection}
%Selection cuts, live-time of each detector, usage of three hexagons
%The data set extends from January 1, 2018, shortly after the completion of the SD-433 deployment campaign, 

The photon search strategy designed for this work relies on the SD-433 data to measure the air showers, providing information on the primary energy and shower geometry, and on the direct measurement of the air-shower muon content, as recorded by the UMD stations. To ensure a sufficient sampling of the air-shower muon content, leading to a suitable discrimination power as discussed in \cref{sec:application}, we select events acquired by the three highlighted hexagons of \cref{fig:UMD433}, left. The data for this analysis were recorded from December 17, 2020, when the central UMD stations in the three employed hexagons were commissioned, to March 31, 2022, before the upgrade of the $19$ stations with new electronics~\citep{Anastasi2022}.

The flux of events acquired by the SD hexagons is expected to remain constant, apart from a negligible seasonal modulation. As the number of detected events follows Poisson statistics, identifying unstable periods in the data acquisition of the SD-433 involves searching for time intervals between consecutive events incompatible with the Poisson expectation\footnote{For example, consecutive events separated more than \unit[1.1]{hours} when the seven hexagons of the array are operative are not compatible with a Poisson probability at a $99\%$ confidence level and hence identified as an unstable period.}. The same methodology is employed for the UMD modules, based on the rate of 6T5 events acquired by the SD-433 during which a UMD module is registered as active. Unstable periods for each UMD module are identified by analyzing the time intervals between consecutive appearances in the data. After subtracting unstable periods for both detection systems based on the arrival time of 6T5 events, the final data set is composed of 5T5 and 6T5 events recorded during $\unit[{\sim}15.5]{months}$, equivalent to $\unit[8]{months}$ of data acquisition with the three hexagons operating under ideal conditions. The final data set is composed of 15,919 events with energies above \unit[$10^{16.7}$]{eV} and zenith angles up to \mbox{$\unit[52]{^\circ}$}. The energy threshold is derived with the Monte Carlo study presented in \cref{sec:PES}, while the zenith angle range is discussed in \cref{sec:triggerEff}. One of these events is shown in \cref{fig:UMD433}, right, where the LDF and energy calibration have been performed using the reconstruction used for the bulk of SD-433 events~\citep{BrichettoICRC2023}.

%Since the event rate of a module is intrinsically linked to its location within the array, only the number of active SD-433 hexagons containing its associated SD station is considered.

%\begin{wrapfigure}[21]{O}{0.5\textwidth}
%\begin{figure}[!tb]
%\includegraphics[width=0.49\textwidth]{./images/auger_210577851800__sd_62300973__md_3275}
%\caption{One of the events acquired on February 27, 2021, and composing the selected data set. The reconstructed energy is \unit[$(4.1\pm0.1)\times10^{17}$]{eV}, obtained with a dedicated data energy scale~\citep{BrichettoICRC2023}, with a zenith angle of \unit[$(26.3\pm0.3)$]{$^\circ$}.}
%\label{fig:exampleEvent}
%\end{figure}
%\end{wrapfigure}

\section{Energy assignment of photon and proton events}
\label{sec:MCRec}

Unlike hadronic-initiated air showers~\citep{Collaboration2021g}, a data-driven energy calibration for photons is impractical because no photons have yet been detected above a few \unit[$\sim10^{15}$]{eV}. Furthermore, given the depleted muon content in photon-initiated showers, a data-driven energy calibration would overestimate the photon energy because the SD is sensitive to muons. Stated otherwise, a photon and a hadronic primary of the same energy do not generate the same average signal in the WCDs. Consequently, simulated air showers with a sufficiently large probability to generate an event in the SD array  are selected in \cref{sec:triggerEff} before being employed to derive a dedicated energy calibration for photon-initiated events\footnote{The bias introduced by the muon deficit in hadronic models is not expected to significantly affect an energy calibration based on photon-initiated events given the minimal muonic content in such showers.}, as described in \cref{sec:photonCal}.

The discrimination method is designed using protons as the only hadronic species, because proton-initiated air showers have a muon content most similar to photon-initiated showers among the hadronic primaries present in the cosmic-ray flux. The energy of simulated proton events is underestimated when employing a data-driven energy calibration, since current high-energy hadronic interaction models appear to underestimate the number of muons produced in air showers~\citep{AMIGAFAL2019,EASMSU2019}. Therefore, a Monte Carlo-based energy assignment is developed in \cref{sec:protonCal}. This approach provides the most conservative background estimation, as discussed in \cref{sec:discrimination}.

Because the nature of the primary cosmic rays is unknown in the data, one cannot apply separate energy calibrations for hadron- and photon-initiated showers. A unified energy scale is thus essential for accurately comparing events initiated by different primary species. To tackle this problem, the photon-equivalent energy scale is developed in \cref{sec:PES}.

%The energy reconstruction relies on the shower size, adjusted for atmospheric attenuation. This calibration is done based on simulations and separately for photons and protons.

\subsection{Trigger efficiency of photon and proton primaries}
\label{sec:triggerEff}

%Simulations are employed to design the event reconstruction of the energy and the discrimination method.

The probability of a shower to trigger the array depends on the characteristics of the primary particle. We quantify this probability, the trigger efficiency, using air-shower simulations. The simulation library covers an energy range between $\unit[10^{16}]{eV}$ and $\unit[10^{17.5}]{eV}$, which follows an $E^{-1}$ distribution and zenith angles uniformly distributed in $\sin^2\theta$ up to \unit[60]{$^\circ$}. It contains 15,000 air showers for each primary, produced using CORSIKA v7.6400~\citep{D.Heck1998} with EPOS-LHC~\citep{Pierog2015} and Fluka2011.2x~\citep{Ballarini2005} as the high- and low-energy hadronic interaction models. The transition energy between these two regimes is set at \unit[80]{GeV}. To manage the computational load, we employ a thinning algorithm as outlined in~\citep{Kobal2001} with a thinning threshold of $10^{-6}$. Subsequently, the particle distributions at ground level are unthinned at the stage of the detector simulation as described in~\citep{Billoir2008}. Each air-shower core is randomly positioned ten times around the central station of the array and the detector response is simulated using the Auger \Offline framework~\citep{Santos2023,Offline2007}.

\begin{figure}[!tb]
\centering
\includegraphics[width=0.49\textwidth]{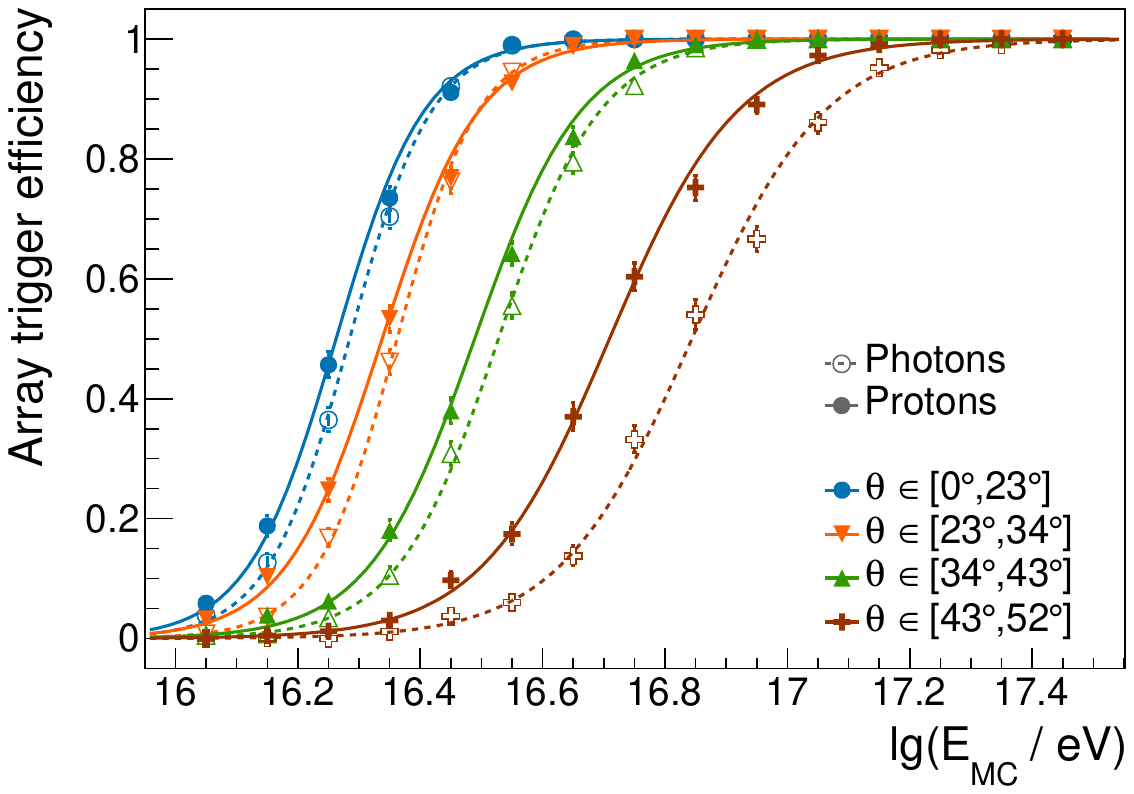}
\caption{The trigger efficiency as a function of the simulated energy for photon and proton primaries, depicted by empty and filled markers respectively, and different zenith angle intervals. Lines correspond to sigmoid functions fitted via a maximum likelihood method.}
\label{fig:TeffE}
\end{figure}

\Cref{fig:TeffE} depicts the trigger efficiency for the simulated photon and proton events. Air showers initiated by hadronic primaries, such as protons, reach the ground with a significantly more prominent muonic component than those initiated by photon primaries due to the decay of charged pions and kaons produced during hadronic interactions in the atmosphere. This increased muonic content is crucial for trigger generation, especially as the zenith angle increases and atmospheric attenuation of the electromagnetic particles becomes more significant. The trigger efficiency for proton-initiated air showers is thus consistently higher than that for photon events, with the difference becoming more pronounced at larger zenith angles.

A sigmoid model, represented by solid lines, describes the trigger efficiency based on the simulated energy, $E_\text{MC}$, and zenith angle, $\theta$, with parameters estimated using a maximum likelihood method. Events generated by air showers with an expected trigger efficiency greater than $90\%$ and a zenith angle smaller than \unit[52]{$^\circ$} are selected for the Monte Carlo-driven analyses throughout this article since the trigger efficiency for photon showers with more inclined directions is negligible. Around 59,000 (64,000) photon (proton) events survive the selection cuts, representing nearly $40\%$ of the total simulated events.

%\begin{wrapfigure}[20]{I}{0.5\textwidth}
%\end{wrapfigure}

\subsection{Photon-initiated events}
\label{sec:photonCal}

%\begin{wrapfigure}[16]{I}{0.5\textwidth}

%\end{wrapfigure}

% Side-by-side figures 
% \begin{figure}
% \begin{minipage}[c]{0.4\linewidth}
% \includegraphics[width=\linewidth]{./images/UMD433}
% \caption{A schematic view of the SD-433 array. Solid black positions are equipped with buried muon detectors. Events acquired by the three highlighted hexagons are considered for the photon search.\label{fig:UMD433}}
% \end{minipage}
% \hfill
% \begin{minipage}[c]{0.5\linewidth}
% \includegraphics[width=\linewidth]{./images/TeffE_fits_v3}
% \caption{The trigger efficiency as a function of the simulated energy for photon primaries and different zenith angle intervals. The efficiency is $85\%$ ($90\%$) at (for energies larger than) \unit[$10^{16.7}$]{eV}, as shown by dashed line. \label{fig:TeffE}}
% \end{minipage}%
% \end{figure}
%Solid lines correspond to the sigmoid functions fitted via a maximum likelihood method. Events with photon energies above $\unit[10^{16.7}]{eV}$, represented by the dashed line, are employed in this work.

%\subsection{Lateral distribution function for photon events}
%\label{sec:LDF}

%[Calibration for photon primaries] Direct energy calibration for photon primaries (CIC and energy parameters). Energy bias and resolution.

As described in \cref{sec:detectorsAndData}, the primary energy is calibrated using the signal interpolated at \unit[250]{m} from the shower axis, $S(250)$. This value is determined by evaluating a parametrized LDF fitted to best match the observed signals in each triggered SD station. To achieve an accurate estimate of $S(250)$, the slope parameter of the LDF is parametrized as detailed in \cref{sec:photonLDF}. We utilize only one of the ten realizations of simulated events produced for each air shower to account for shower-to-shower fluctuations in the energy calibration. A power-law model is used to describe the relationship between $S(250)$ and $E_\text{MC}$~\citep{Spectrum2020}:

%On the one hand, this allows us to properly include the shower-to-shower fluctuations. On the other hand, the impact of a possible bias due to multiple simulations of very deep showers, especially those that reach the ground before their maximum development, is reduced.
%The LDF reconstruction for simulated photon events with a trigger efficiency above $90\%$ is carried out using the model of \cref{eq:LDF} and the slope parametrization \mbox{from~\cref{eq:beta}}. 
% Inspired by \cref{fig:S250EMC}, left, we propose a

\begin{equation}
\label{eq:S250E}
\frac{S(250)}{g(\theta)} = \left( \frac{E_\text{MC}}{10^{17}\,\text{eV}} \right)^{\alpha(\theta)}
\end{equation}

\begin{figure}[!tb]
\includegraphics[width=0.49\textwidth]{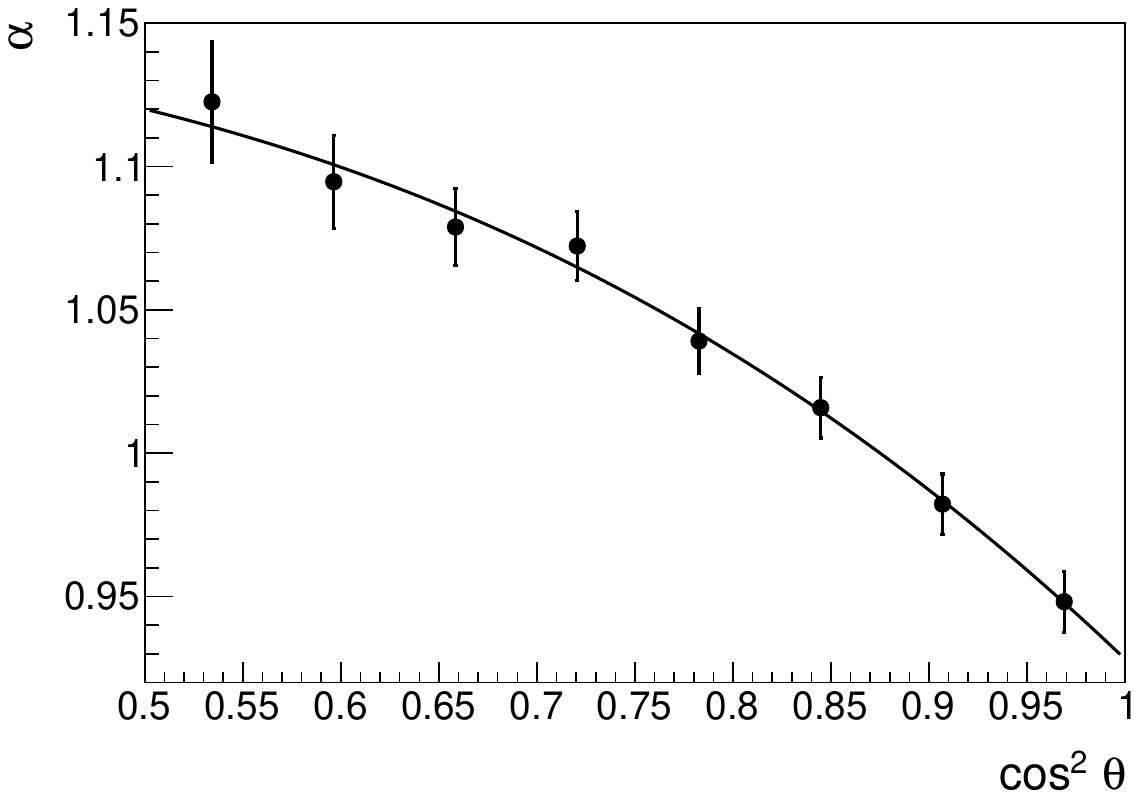} \includegraphics[width=0.49\textwidth]{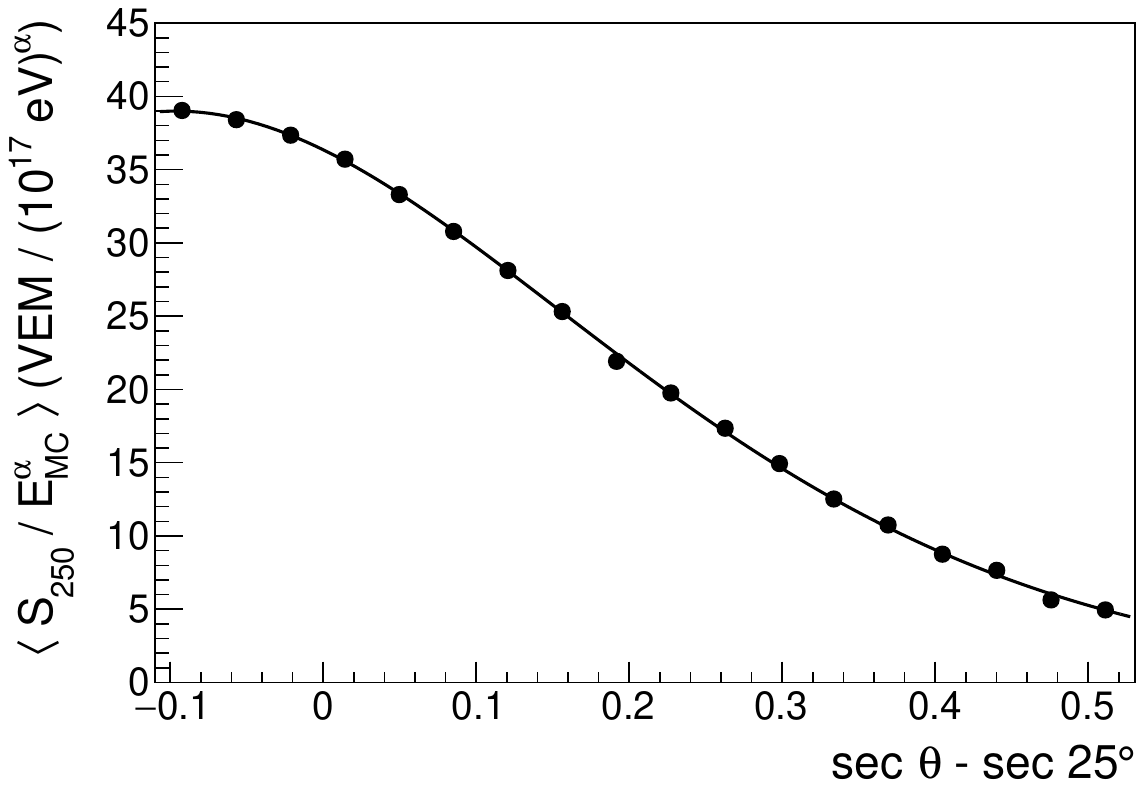}
\caption{Left: The power-law index from \cref{eq:S250E} as a function of zenith angle. The solid line represents the model of \cref{eq:alphaCos2}. Right: The average ratio between shower size and primary energy raised to the power of $\alpha$, in terms of the zenith angle. The fitted model of \cref{eq:GH} is represented as the solid line.}
\label{fig:photonScale}
\end{figure}

\noindent The power-law index, $\alpha$, is found for the subset of events in each of eight zenith-angle bins employing a $\chi^2$ minimization. As shown in \cref{fig:photonScale}, left, $\alpha$ is mildly dependent on the zenith angle with a relationship that can be modeled as:

\begin{equation}
\label{eq:alphaCos2}
\alpha(\theta) = \alpha_0 \times \left( 1 + \alpha_1\times (\cos^2\theta)^{\alpha_2} \right)
\end{equation}

\begin{figure}[htbp]
\centering
\begin{minipage}{0.48\textwidth}
\centering
\includegraphics[width=\textwidth]{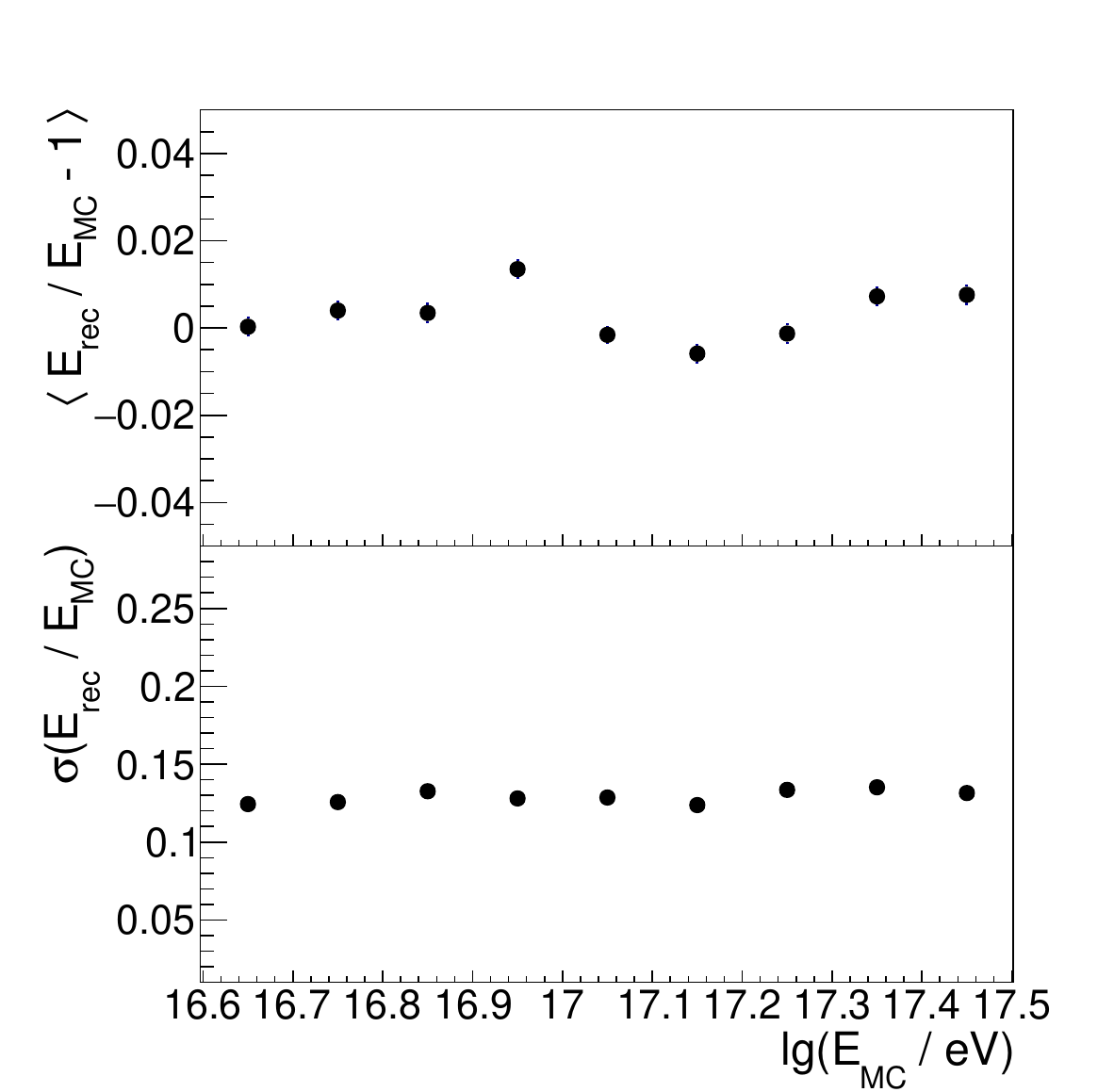}
\caption{The bias and resolution of the energy calibration for photon events defined by \cref{eq:S250E}.}
\label{fig:BiasResoEMC_ph}
\end{minipage}\hfill
\begin{minipage}{0.48\textwidth}
\centering
\vspace{0.92cm}
\begin{table}[H]
\centering
\begin{tabular}{c c}
\toprule
{Parameter} & {Value} \\ 
\midrule
\midrule
{$\alpha_0$} & {$1.15\pm0.04$} \\
{$\alpha_1$} & {$-0.192\pm0.025$} \\
{$\alpha_2$} & {$2.96\pm1.17$} \\
\midrule
{$g_0$} & {$38.9\pm0.1\,$VEM} \\
{$g_1$} & {$0.623\pm0.053$} \\
{$g_2$} & {$-(9.64\pm0.69)\times10^{-2}$} \\
{$g_3$} & {$(9.05\pm0.48)\times10^{-2}$} \\
\bottomrule
\vspace{0.92cm}
\end{tabular}
\caption{\label{tab:AttPars} The parameters modeling the attenuation of the shower size and the energy calibration, defined by \cref{eq:alphaCos2,eq:GH}.}
\end{table}
\end{minipage}
\end{figure}

\noindent The three free parameters are listed in \cref{tab:AttPars}. The power-law index $\alpha$ is slightly below unity for vertical events due to the higher likelihood of these showers arriving at the observation level before reaching their maximum development, especially at the highest energies, compared to non-vertical events. The ratio $S(250)/E^\alpha$ decreases with the zenith angle, as shown in \mbox{\cref{fig:photonScale}}, right, reflecting the atmospheric attenuation of electromagnetic air showers. Based on the universality of the electromagnetic longitudinal development, the signal-to-energy ratio can be described using a Gaisser-Hillas function~\citep{Gaisser1977_1}:

\begin{equation}
\label{eq:GH}
g(\theta) = g_0 \times \left(1+\frac{x-g_2}{g_1}\right)^{g_1/g_3} \times \exp\left(-\frac{x-g_2}{g_3}\right),\,\text{where}\, x = \sec\theta - \sec\unit[25]{^\circ}
\end{equation}

\noindent The free parameters, listed in \cref{tab:AttPars}, are estimated via a $\chi^2$ minimization to the average ratio. Particularly, the parameter $g_0=\unit[(38.9\pm0.1)]{VEM}$ represents the expected signal measured by the SD-433 at \unit[250]{m} from the axis of a shower initiated by a photon primary of \unit[$10^{17}$]{eV} with a zenith angle of \unit[25]{$^\circ$}. 

The calibration performance is evaluated by examining the relative difference between the reconstructed energy, $E_\text{rec}$, and the simulated energy, $E_\text{MC}$. This method reconstructs the photon energy with a bias of less than $2\%$, as shown in \cref{fig:BiasResoEMC_ph}, top. The resolution of the calibration is defined as the standard deviation of the distribution of the relative difference between $E_\text{rec}$ and $E_\text{MC}$. As shown in \cref{fig:BiasResoEMC_ph}, bottom, the zenith-integrated resolution remains nearly constant at around $12\%$.
%, slightly increasing as the primary energy increases. This is due to the angular dependence of the resolution. Indeed, events with $\theta\in[\unit[40]{^\circ},\unit[52]{^\circ}]$ have a resolution between of $30\%$ and $20\%$, improving with the primary energy, while those with $\theta\in[\unit[0]{^\circ},\unit[27]{^\circ}]$ are reconstructed with an energy resolution better than $10\%$.

%\begin{wrapfigure}[19]{I}{0.5\textwidth}
%\begin{figure}[!tb]
%\centering
%\includegraphics[width=0.49\textwidth]{./images/BiasEMC_ph_PhotonScale.pdf} \includegraphics[width=0.49\textwidth]{./images/ResoEMC_v4.pdf}
%\includegraphics[width=0.49\textwidth,trim=0cm 0cm 0cm 1.5cm]{./images/BiasResoEMC_ph_PhotonScale.pdf}
%\caption{The bias and resolution of the energy calibration for photon events defined by \cref{eq:S250E}.}
%\label{fig:BiasResoEMC_ph}
%\end{figure}
%\end{wrapfigure}

%\begin{table}[!tb]
%\begin{wraptable}{R}{0.5\textwidth}
%\begin{center}
%\begin{tabular}{c c}
%\toprule
%{Parameter} & {Value} \\ 
%\midrule
%\midrule
%{$\alpha_0$} & {$1.15\pm0.04$} \\
%{$\alpha_1$} & {$-0.192\pm0.025$} \\
%{$\alpha_2$} & {$2.96\pm1.17$} \\
%\midrule
%{$g_0$} & {$38.9\pm0.1\,$VEM} \\
%{$g_1$} & {$0.623\pm0.053$} \\
%{$g_2$} & {$-(9.64\pm0.69)\times10^{-2}$} \\
%{$g_3$} & {$(9.05\pm0.48)\times10^{-2}$} \\
%\bottomrule
%\end{tabular}
%\end{center}
%\caption{\label{tab:AttPars} The parameters modeling the attenuation of the shower size and the energy calibration, defined by \cref{eq:alphaCos2,eq:GH}.}
%\end{wraptable}
%\end{table}

%As shown in \cref{fig:biasE}, the energy bias of the calibration is negligible when integrating all zenith angles, while the bias is within $5\%$ for events with $\theta>34^\circ$. A quantification of the resolution is attained by studying the relative energy bias for each energy and zenith bin. Each distribution is described with a Gaussian function and the standard deviation is designated as the resolution estimator. The energy resolution is at the level of $10\%$ ($20\%$) for events with $\theta<34^\circ$ ($34^\circ<\theta<43^\circ$) with a very small dependence on the energy, as shown in \cref{fig:resoE}. The resolution drops to $\sim25\%$ for $\theta>43^\circ$. The zenith-integrated resolution is mostly constant at $\sim12\%$ with a mild increasing trend with the primary energy.

\subsection{Proton-initiated events}
\label{sec:protonCal}

%As discussed above in the context of photonic events,
 
%Although there is a data-driven energy calibration tailored for SD-433 events~\citep{BrichettoICRC2023}, a dedicated energy calibration for proton-simulated events is desired to account for a possible underestimation of the number of muons in current hadronic interaction models employed in air-shower simulation~\citep{EASMSU2019}.
%The trigger efficiency for proton primaries must be studied before designing a dedicated energy calibration. The trigger efficiency for proton events can be estimated using the previously described simulated library, applying a similar maximum likelihood method with a sigmoid model. 

%\begin{wrapfigure}[16]{I}{0.5\textwidth}
%\begin{figure}[!tb]
%\centering
%\includegraphics[width=0.5\textwidth]{./images/TeffE_fits_proton_v2.pdf}
%\caption{The trigger efficiency as a function of the simulated energy for proton primaries and different zenith angle intervals. Solid lines correspond to the sigmoid functions fitted via a maximum likelihood method, while dashed lines are those obtained for photon events as displayed in \cref{fig:TeffE}.}
%\label{fig:TeffE_pr}
%\end{figure}
%\end{wrapfigure}

%The proton-initiated events generated by air showers with an expected trigger efficiency above $90\%$ are selected for the energy calibration described in this subsection. Around 64,000 events survive the selection cut, representing nearly $42\%$ of the total simulated proton events.
This subsection details the energy assignment of simulated proton events, focusing on the relationship between $S(250)$ and $E_\text{MC}$. As discussed above in the context of photon events, a single realization per air shower is used to properly account for shower-to-shower fluctuations. The LDF slope parametrized for the bulk of the SD-433 data~\citep{BrichettoICRC2023} is utilized for the simulated proton events since it is suitable to describe hadronic-initiated events. A power-law is used to describe the correlation between $S(250)$ and $E_\text{MC}$
%the photon energy calibration, i.e.,\

\begin{equation}
\label{eq:S250E_pr}
\frac{S(250)}{g'(\theta)} = \left( \frac{E_\text{MC}}{10^{17}\,\text{eV}} \right)^{\alpha'}
\end{equation}
%S_{250}\left(E_\text{MC},\theta|\overrightarrow{p}\right) / \text{VEM} = g'(\theta) \times \left(\frac{E_\text{MC}}{E_0}\right)^{\alpha'}

\noindent in analogy with \cref{eq:S250E}. The attenuation function model for proton events, inspired by recent analyses with the SD-433 data~\citep{Silli2021_2,BrichettoICRC2023}, is
%where the unknown parameters are jointly defined as $\overrightarrow{p}$.

\begin{equation}
\label{eq:fatt_pr}
g'(\theta) = a' \times \left(1 + b' \times x + c' \times x^2 + d' \times x^3 \right),\,\text{where}\, x = \cos^2\theta - \cos^2\unit[30]{^\circ}
\end{equation}

\noindent We select a reference zenith angle of \unit[30]{$^\circ$}, as it corresponds to the median zenith angle of the simulated proton events, ensuring a representative selection. A combined fit of the attenuation function, $g'(\theta)$, and the power-law index, $\alpha'$, to the $S(250)$ is conducted. In this process, the likelihood of observing the samples of $S(250)$ is maximized by tuning the mentioned free parameters assuming an underlying Gaussian probability density function describing the theoretical shower size. The standard deviation of the predicted shower size is taken as the uncertainty coming from the event reconstruction. The estimated values for the parameters involved in \cref{eq:S250E_pr,eq:fatt_pr} are summarized in \cref{tab:ECalib_pr}.

%\footnote{The reference zenith angle for the proton case is larger than for the photon case. This is due to the increased trigger efficiency for protons at higher zenith angles.}

%, ensuring that the calibration model is well-centered around a representative event
%, the normalization energy, $E_0$,
\begin{table}[!tb]
\begin{center}
\begin{tabular}{c c c c c}
\toprule
{$a'$ (VEM)} & {$b'$} & {$c'$} & {$d'$} & {$\alpha'$} \\
\midrule
{$(35.2\pm0.1)$} & {$(1.88\pm0.01)$} & {$(-1.74\pm0.01)$} & {$(-3.45\pm0.05)$} & {$(1.02\pm0.01)$} \\
\bottomrule
\end{tabular}
\end{center}
\caption{\label{tab:ECalib_pr} The parameters in \cref{eq:S250E_pr,eq:fatt_pr} estimated through the maximum likelihood method on the $S(250)$ for proton-initiated events.}
\end{table}
%& {$E_0 / \unit[10^{15}]{eV}$}
%& {$(3.05\pm0.01)$}

\begin{wrapfigure}[21]{O}{0.5\textwidth}
\includegraphics[width=0.49\textwidth]{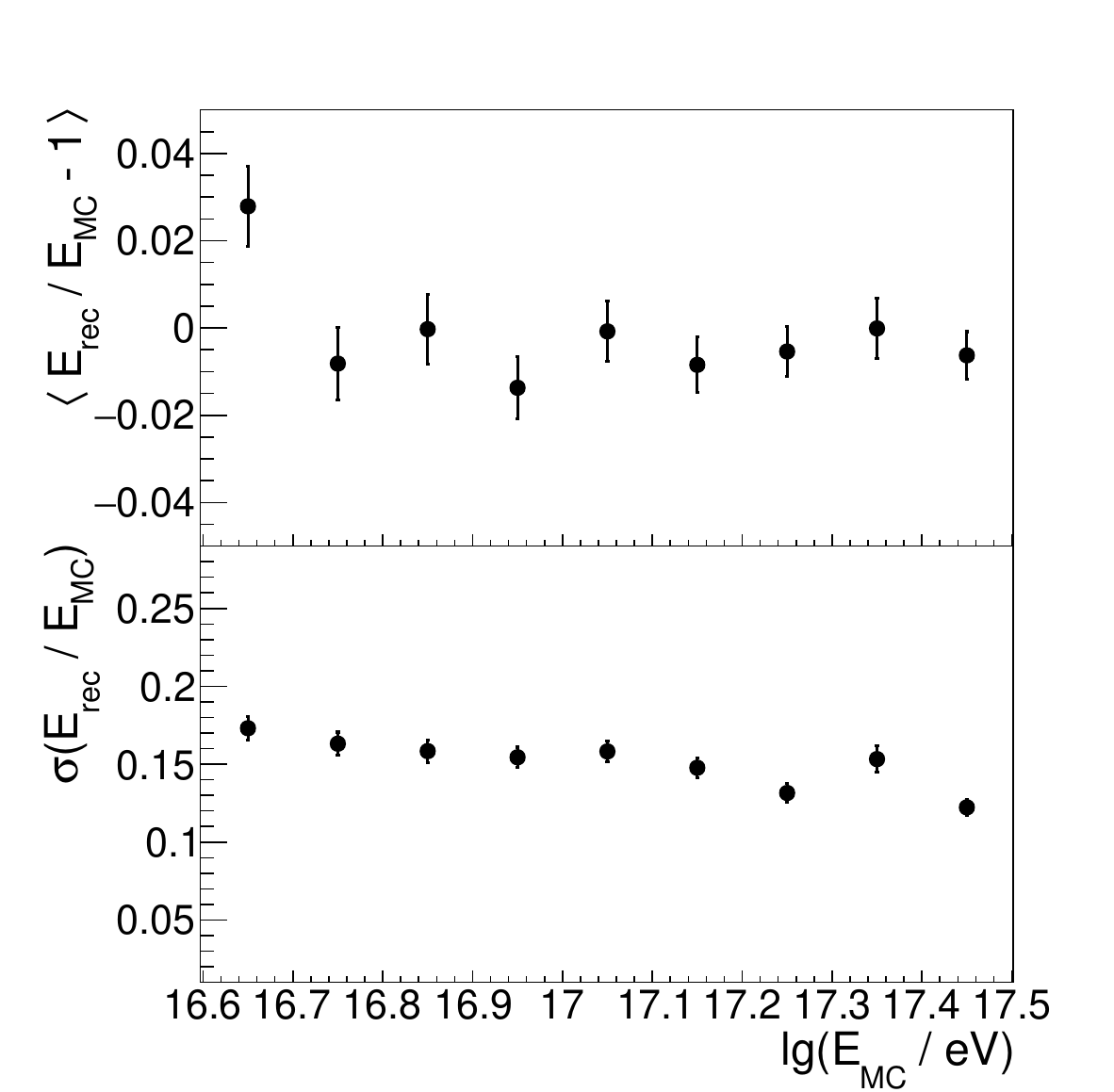}
\caption{The bias (top) and resolution (bottom) of the energy calibration for proton events defined by \cref{eq:S250E_pr}.}
\label{fig:ECalProtons}
\end{wrapfigure}

The bias and resolution provide insight into the reliability and precision of the calibration. As shown in the top panel of \cref{fig:ECalProtons}, the reconstructed energy exhibits a bias below $2\%$ for energies above \unit[$\sim10^{16.7}$]{eV}, while the resolution follows the expected trend with the primary energy, improving from $18\%$ to $13\%$ across the energy range under study, as presented in the bottom panel of \cref{fig:ECalProtons}.

%The former represents the systematic deviation of the reconstructed energy, $E_\text{rec}$, from the true simulated energy, $E_\text{MC}$, while the latter reflects the spread of the reconstructed energy distribution around the true energy.

\subsection{The photon-equivalent energy scale}
\label{sec:PES}

%[Photonic energy scale] Explain why we need a single energy scale for sims and data.

\begin{figure}[!tb]
\includegraphics[width=0.49\textwidth]{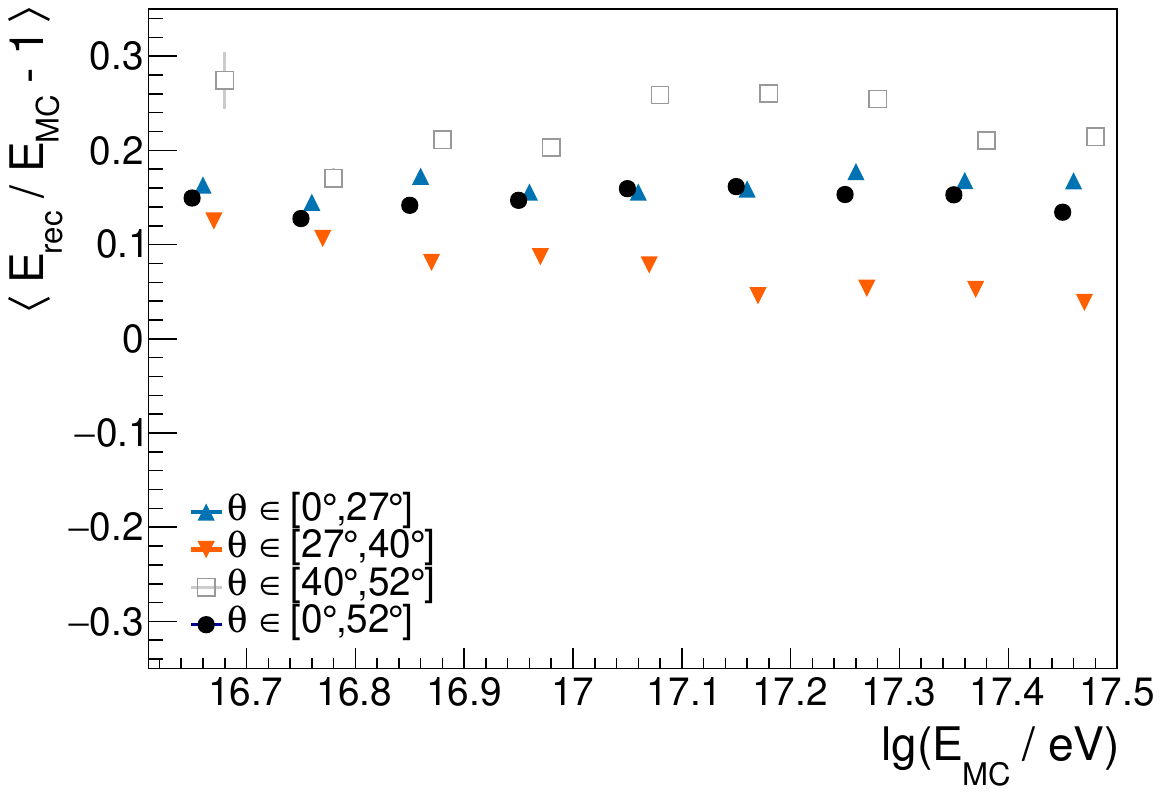} \includegraphics[width=0.49\textwidth]{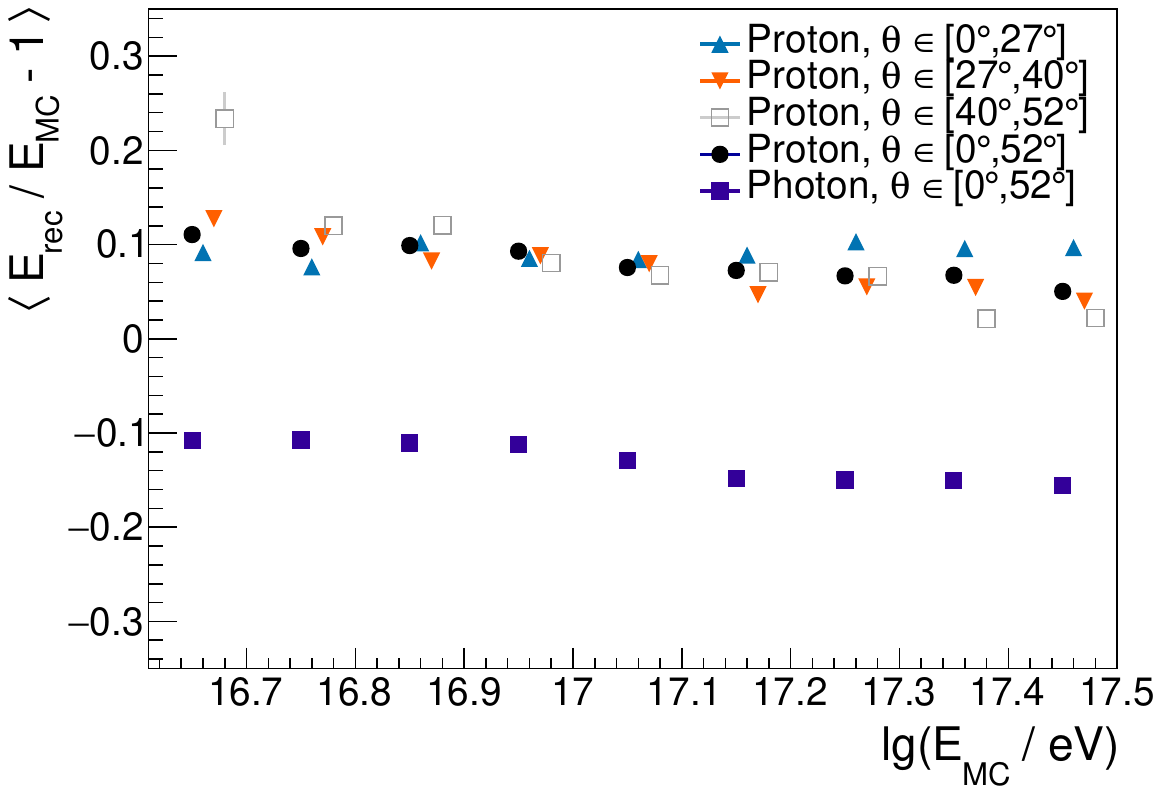}
%\subfigure[\label{fig:resoE}]{\includegraphics[width=0.49\textwidth]{./images/ResoEMC}}
\caption{The relative difference between the reconstructed and simulated energy in terms of the latter. Left: Events are initiated by proton primaries, where the energy is assigned using the procedure described in \cref{sec:photonCal}. Right: Events are initiated by photon and proton primaries, where the energy is assigned using the photon-equivalent energy scale (see text for details).}
\label{fig:photonScaleAndPES}
\end{figure}

In the case of simulated proton events, the energy assignment tailored for photon primaries described in \cref{sec:photonCal} leads to a systematic overestimation of the reconstructed energy by $\sim15\%$, as shown by the black markers in \cref{fig:photonScaleAndPES}, left. This is due to the muonic component, larger in hadronic showers than in electromagnetic ones, in particular at \unit[250]{m} from the shower axis. Furthermore, because proton and photon showers attenuate differently in the atmosphere, one can see in the figure a clear dispersion in the reconstructed energy across angular bins. To account for the different attenuation of hadron- and photon-initiated air showers, we define the photon-equivalent energy scale, $E_{\gamma,\text{eq}}$. This scale is constructed using the attenuation curve derived from simulated proton events, i.e.,\ from \cref{eq:fatt_pr}, and the energy power-law index obtained for photon events, i.e.,\ from \cref{eq:S250E}. Thus a reconstructed event with a shower size $S(250)$ can be assigned an energy $E_{\gamma,\text{eq}}$ independently of the primary species and for both simulations and data.
%, in analogy with \cref{eq:S250E}
%, i.e.,\ the pivot energy and power-law index

The application of this energy scale to simulated proton events results in a bias of $10\%$ at $\unit[10^{16.7}]{eV}$ decreasing to $5\%$ at the highest energies, as shown in \cref{fig:photonScaleAndPES}, right. Importantly, this bias shows no significant angular dependence, making $E_{\gamma,\text{eq}}$ a robust scale across different zenith angles. On the other hand, photon-initiated events, when analyzed using this scale, are assigned underestimated energies. The bias changes from $-10\%$ to $-15\%$ in the energy range of interest, as displayed in \cref{fig:photonScaleAndPES}, right. Consequently, in a given $E_{\gamma,\text{eq}}$ bin, simulated photon events are mixed with proton events that have true lower energies, and thus lower muon content. This mixing leads to a conservative estimation of the separation between hadron- and photon-initiated events based on the muon content that scales with the primary energy. Therefore, the discrimination method described in the next section is tailored and optimized in such a conservative scenario.

%\begin{equation}
%\label{eq:photonicScale}
%S(250)^\text{pr} = g^\text{pr}(\theta) \times E_{\gamma,\text{eq}}^\alpha
%\end{equation}

%\begin{figure}[!tb]
%\subfigure[\label{fig:biasWithPhotonEcal}]{\includegraphics[width=0.49\textwidth]{./images/BiasEMC_protons}}
%\subfigure[\label{fig:biasWithPhotonicScale}]{\includegraphics[width=0.49\textwidth]{./images/BiasEMC_protons_pes}}
%\caption{The energy bias for proton events reconstructed with the proton LDF parametrization using (a) the attenuation curve and energy calibration developed for photon-induced events and (b) the proton attenuation curve and the photon energy calibration, i.e. the photon-equivalent energy scale.}
%\end{figure}

%\begin{wrapfigure}[14]{i}{0.5\textwidth}
%\begin{figure}[!tb]
%\centering
%\includegraphics[width=0.49\textwidth]{./images/Fatt_phpr}
%\caption{The attenuation functions intervening in the energy calibration for photon and proton primaries.}
%\label{fig:fattPhPr}
%\end{figure}
%\end{wrapfigure}

% reconstructed using the energy assignment photon attenuation function and calibration slope (i.e.,\ \cref{eq:S250E})

\section{Discrimination between photon and proton events}
\label{sec:discrimination}

%[Underlying physics] Introduce observable and describe its features
The search for primary photons in the cosmic-ray flux is a classification problem of a tiny signal in the overwhelmingly dominant hadronic background. The muon density is a powerful observable for distinguishing between air showers dominated by the electromagnetic component and those with significant hadronic contributions. Using the simulated set described in \cref{sec:triggerEff}, we define in \cref{sec:definitionMb} a muon-based observable that can be related to the mass of the primary cosmic ray. We optimize it to maximize the separation power between photon- and proton-initiated events in \cref{sec:optimizationMb}, where the energy is assigned with the photon-equivalent energy scale, as discussed in \cref{sec:PES}.

%This energy assignment causes proton events with lower simulated energies to migrate to higher energy bins. 
%photonic and hadronic events.The detection of muons at ground level is a well-established method for distinguishing between
%Air-shower muons reaching the ground with kinetic energy higher than $\unit[\sim1]{GeV}\times\sec\theta$ are directly measured by the UMD stations~\citep{DeJesusICRC2023,AMIGAFAL2019}.

\subsection[Muon content estimator, $M_b$]{Muon content estimator, \boldmath{$M_b$}}
%\texorpdfstring{$M_b$}{Mb
\label{sec:definitionMb}

For each event, the muon densities, $\rho_i$, measured at various distances, $r_i$, from the shower axis are combined to define the discrimination observable $M_b$~\citep{GonzalezAstropartPhys2019}:

\begin{equation}
M_b = \lg \left( \sum_{i} \frac{\rho_i}{\rho_\text{pr}} \times \left(\frac{r_i}{r_\text{pr}}\right)^b \right)
\label{eq:Mb}
\end{equation}
%, a weighted sum designed to enhance the separation between electromagnetic-dominated and hadronic-dominated showers

Each measured muon density is normalized by the average muon density, $\rho_\text{pr}$, expected in proton-initiated events at a predefined reference distance, $r_\text{pr}$, and is parametrized in terms of the primary energy and zenith angle. This normalization serves a dual purpose in the discrimination method. First, it offers an initial estimate of the muon content, with hadronic-induced showers typically yielding positive values of $M_b$ and photon-initiated showers producing negative values. Second, by normalizing each muon density by $\rho_\text{pr}$, the energy and angular dependencies of $M_b$ are largely absorbed, while the composition-dependent scaling of the muon densities is preserved. The power-law index, $b$, is selected to maximize the separation power of $M_b$ between photonic and hadronic events, as discussed in~\cref{sec:optimizationMb}.

%\begin{wrapfigure}[20]{I}{0.5\textwidth}
%\includegraphics[width=0.49\textwidth]{./images/Rhor_16.7_16.8_MCData}
%\caption{The average measured muon density in terms of the shower axis for simulated events produced by the quoted primary cosmic rays with simulated energies \mbox{$\lg\left(E_\text{MC}/\text{eV}\right) \in (16.7,16.8)$} and for data events with reconstructed energies in the same nominal interval.}
%\label{fig:RhorMCData}
%\end{wrapfigure}

\begin{figure}[!tb]
\includegraphics[width=0.49\textwidth]{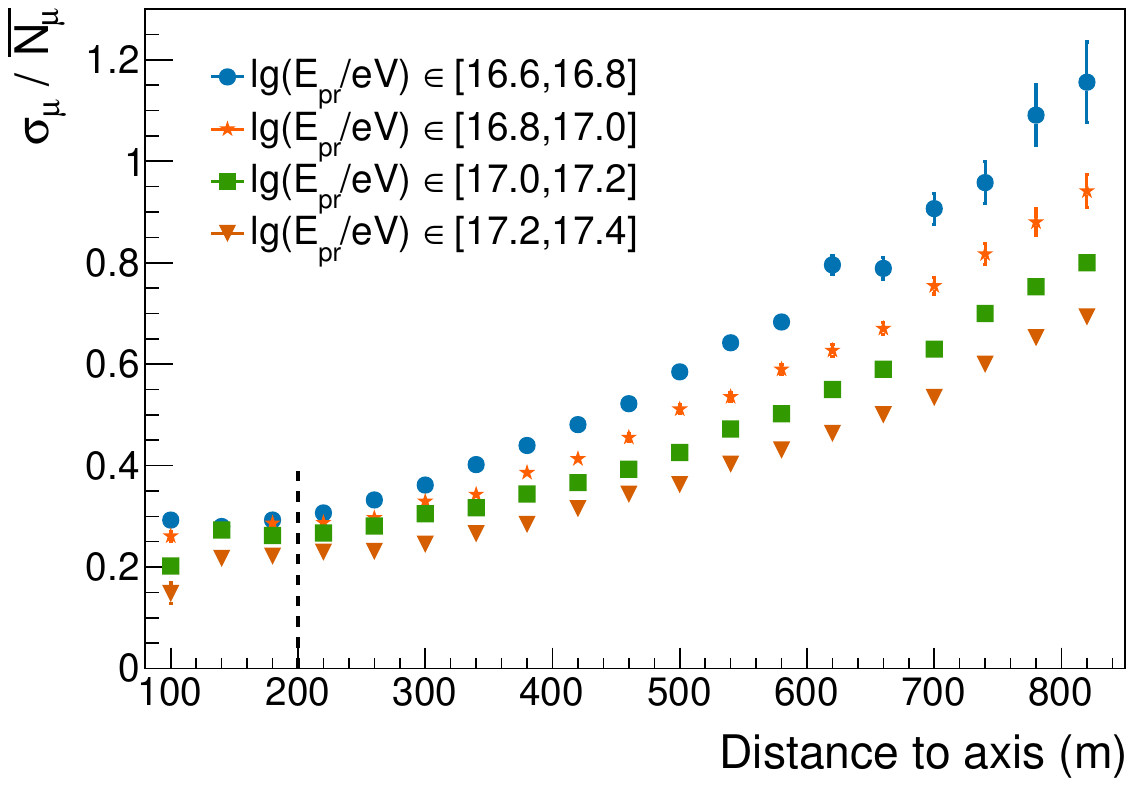}\includegraphics[width=0.49\textwidth]{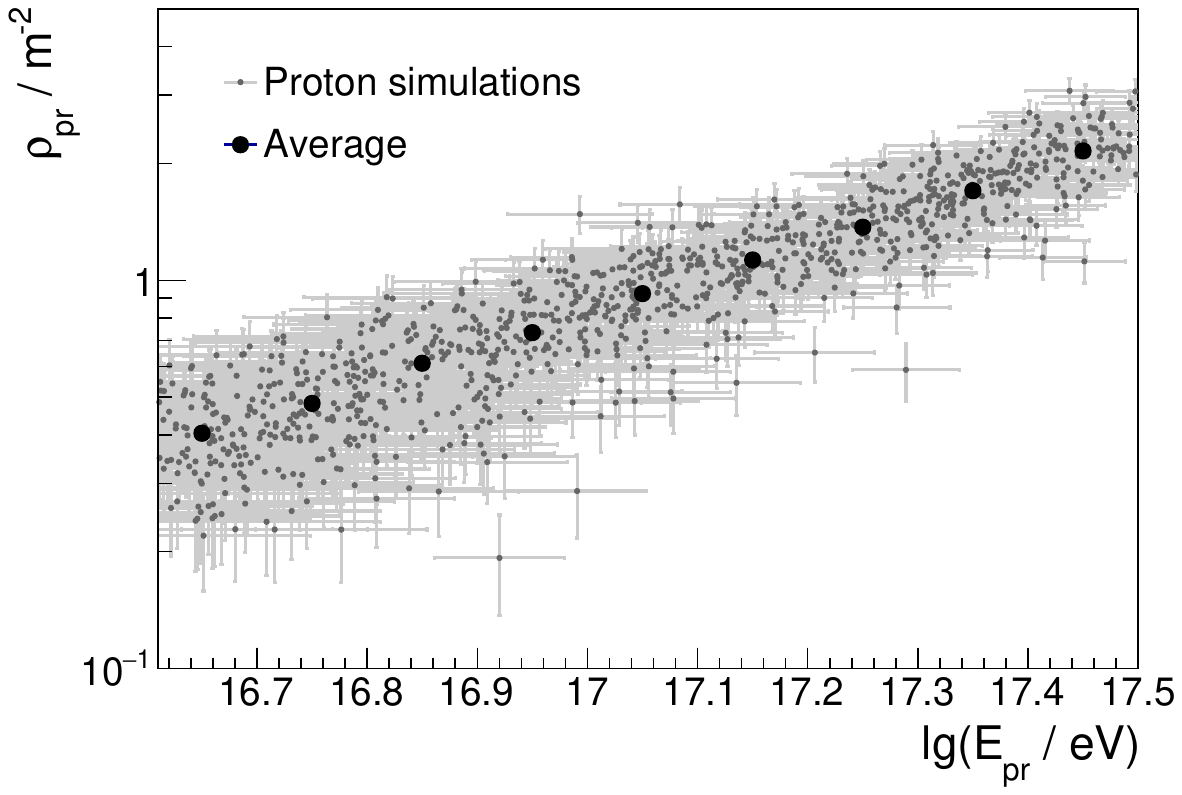}
\caption{Left: The ratio between the standard deviation of the number of impinging muons on UMD stations and its average value as a function of the distance to the shower axis for proton events in the quoted reconstructed energy intervals. The vertical dashed line represents the choice of $r_\text{pr}=\unit[200]{m}$. Right: The muon density at \unit[200]{m} from the shower axis for proton-initiated showers as a function of the reconstructed energy. In both panels, events with $\theta<\unit[23]{^\circ}$ are employed, as an example.}
%the reconstructed energy is obtained as described in \cref{sec:protonCal}, and 
\label{fig:RhoRef}
\end{figure}

The reference distance, $r_\text{pr}$, is chosen to minimize fluctuations in the number of muons intersecting the sensitive area of a UMD station. The average number of muons, $\overline{N_\mu}$, decreases with the distance to the shower axis, following the lateral spread of the shower front. Conversely, the relative fluctuations, $\sigma_\mu/\overline{N_\mu}$, increase with distance, as shown in \cref{fig:RhoRef}, left. These relative fluctuations are influenced by both the shower-to-shower and the Poissonian fluctuations sprouting from the decreasing number of impinging muons~\citep{Kizakke2024}. The relative fluctuations are estimated as the root-mean-square of the $N_\mu$ distributions in distance bins\footnote{At distances below \unit[$\sim$300]{m}, shower-to-shower fluctuations dominate in the energy range under study, exceeding the Poissonian fluctuations by a factor between $2$ and $3$ at distances between \unit[$10^{16.5}$]{eV} and \unit[$10^{17}$]{eV}.}. A plateau in the relative fluctuations is observed up to \unit[$\sim300$]{m}, mildly dependent on the primary energy and zenith angle. Consequently, we select a reference distance of $r_\text{pr}=\unit[200]{m}$, which corresponds to about $25\%$ statistical fluctuations in the number of muons with a minimal variation across the studied energy range.

%For this analysis, simulated proton-initiated events with an expected trigger efficiency above $90\%$ are selected (see \cref{sec:protonCal}).
%Beyond \unit[200]{m}, Poissonian fluctuations become more significant as the number of muons decreases with distance, although this effect is partially compensated with a higher primary energy.
%\footnote{The high-energy nature of the muons measured by the UMD stations results also in a negligible angular dependence on these fluctuations.}. 

%{\textbf{\color{purple}{Parametrization of \text{$\rho_\mu^{ref}$}:}} 
The normalization factor, $\rho_\text{pr}$, is then defined as the average of the muon density in proton-induced events at \unit[200]{m} from the shower axis. It is parameterized in terms of the proton energy, $E_\text{pr}$, and zenith angle using the muon densities measured between \unit[$195$]{m} and \unit[$205$]{m} from the shower axis. The energy dependence of $\rho_\text{pr}$ is illustrated in \cref{fig:RhoRef}, right, and can be described by:
%in the set of simulated proton events
%, like the relation between the shower size and the primary energy
\begin{equation}
\label{eq:RhoRefE}
\rho_\text{pr} = \rho_0(\theta) \times \left(\frac{E_\text{pr}}{\unit[10^{17}]{eV}}\right)^{c(\theta)} 
\end{equation}

\begin{wrapfigure}[20]{O}{0.5\textwidth}
\centering
\includegraphics[width=0.49\textwidth,trim=0cm 0cm 0cm 1.5cm]{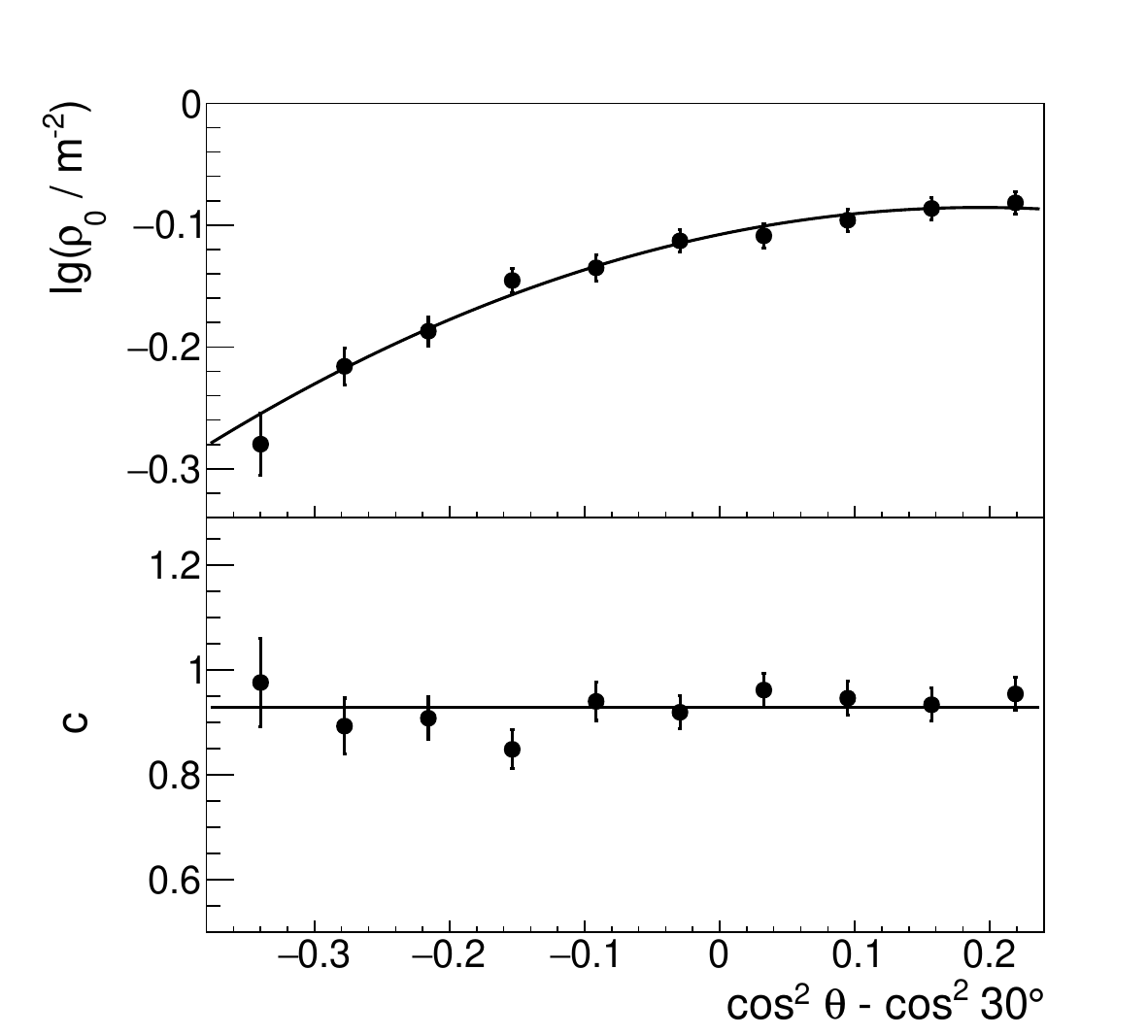}
\caption{The free parameters of \cref{eq:RhoRefE} estimated by $\chi^2$ minimization to the muon densities in angular bins with superimposed fitted quadratic and constant models.}
\label{fig:RhoRefE_rho0c}
\end{wrapfigure}

\noindent The estimated values of the free parameters, $\rho_0$ and $c$, using events in different zenith angle intervals are depicted in \cref{fig:RhoRefE_rho0c}. The normalization, $\rho_0$, decreases with the zenith angle, representing the atmospheric attenuation of the muon component. In contrast, the power-law index, $c$, governs the energy-driven muon production and shows no significant angular dependence. A quadratic dependence on $\cos^2\theta - \cos^{2}\unit[30]{^\circ}$ is proposed to describe $\lg(\rho_0)$, while $c$ is assumed to be a constant. These four free parameters are estimated using a maximum likelihood approach to the unbinned data and presented in \cref{tab:RhoRefML}.
%which is fitted to the simulated data in different angular bins.

%\begin{figure}[!tb]
%\subfigure[\label{fig:RhoRefE_rho0}]{\includegraphics[width=0.5\textwidth]{images/Rho0cos2}}
%\subfigure[\label{fig:RhoRefE_c}]{\includegraphics[width=0.5\textwidth]{images/Ccos2}}
%\caption{The fitted values of the parameters of \cref{eq:RhoRefE} to the unbinned data in terms of $\cos^2\theta - \cos^{2}(30^\circ)$ with superimposed quadratic and constant fits (see text for details).} 
%\end{figure}

\begin{table}[!tb]
\begin{center}
\begin{tabular}{c c c c c}
\toprule
{} & {$\rho_{00}$} & {$\rho_{01}$} & {$\rho_{02}$} & {$c$} \\
\midrule
\midrule
{Mean} & {$-0.108$} & {$0.262$} & {$-0.591$} & {$0.890$} \\
%{Uncertainty} & {$2.13 \cdot 10^{-3}$} & {$9.82 \cdot 10^{-3}$} & {$5.46 \cdot 10^{-2}$} & {$5.63 \cdot 10^{-3}$} \\
{Uncertainty} & {$\phantom{+}0.002$} & {$\phantom{+}0.010$} & {$\phantom{+}0.055$} & {$\phantom{+}0.006$} \\

\bottomrule
\end{tabular}
\end{center}
\caption{\label{tab:RhoRefML} The constant, linear and quadratic coefficients in $\cos^2\theta - \cos^{2}\unit[30]{^\circ}$ describing $\lg(\rho_0(\theta))$ in \cref{eq:RhoRefE}. The last column corresponds to the parameter $c$ present in \cref{eq:RhoRefE}.}
\end{table}

%\subsection{Optimization of \texorpdfstring{$M_b$}{Mb}}
\subsection[Optimization of $M_b$]{Optimization of \boldmath{$M_b$}}
\label{sec:optimizationMb}

%Mb distributions. Definition of signal efficiency and background contamination. Choice of b

\begin{figure}[!tb]
\includegraphics[width=0.5\textwidth]{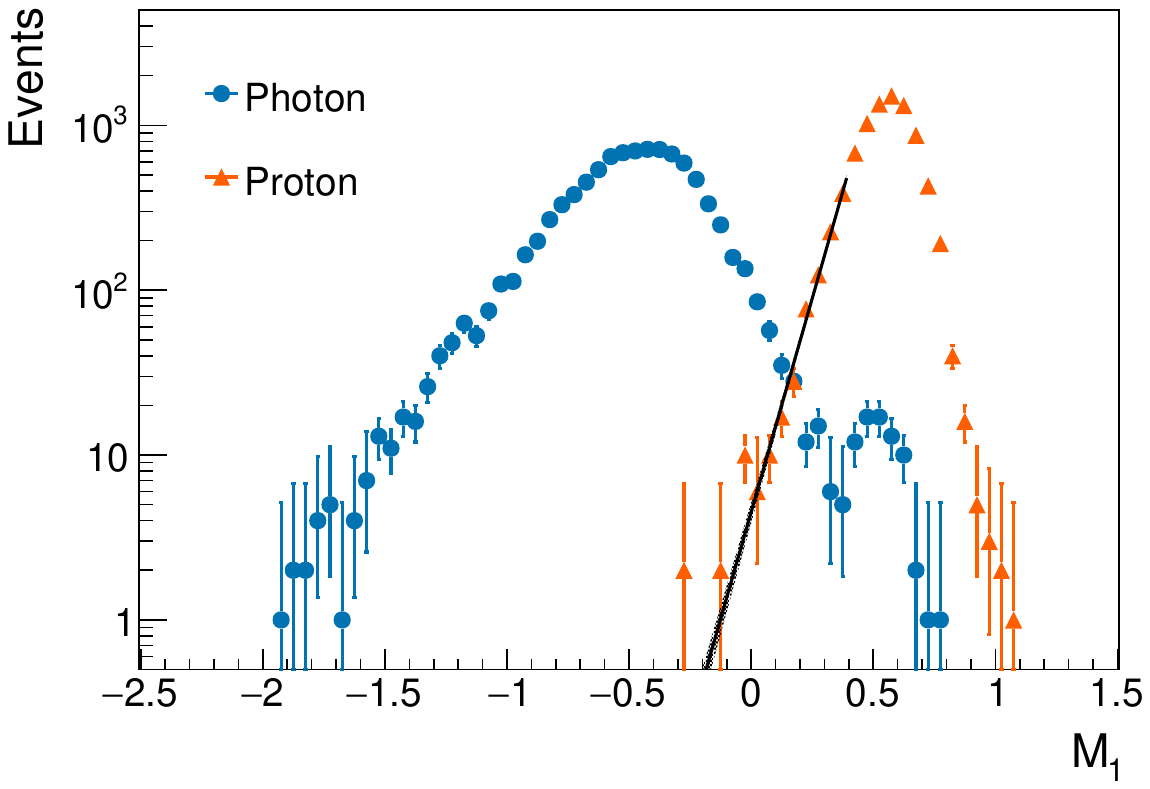}\includegraphics[width=0.5\textwidth]{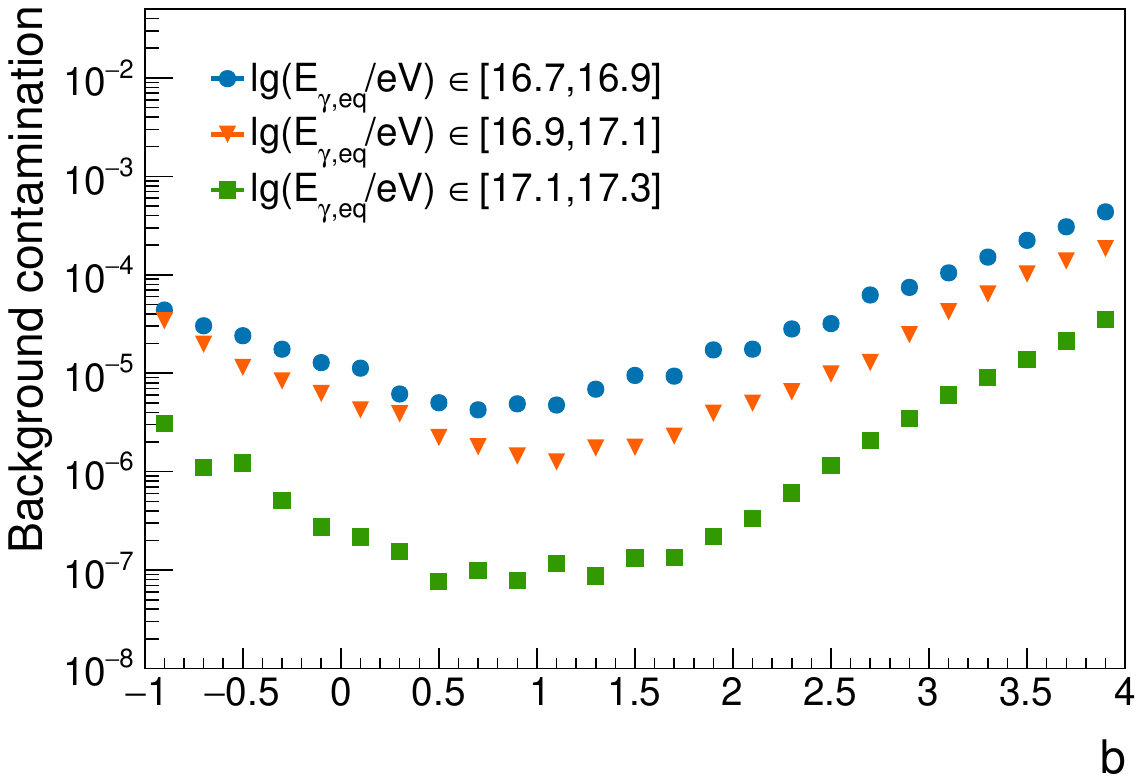}
\caption{Left: The $M_{b=1}$ distributions for simulated photon- and proton-initiated events for energies $\lg\left(E_{\gamma,\text{eq}}/\text{eV}\right) \in (16.7,16.9)$. The uncertainty in bins with fewer than ten entries is the asymmetric confidence belt calculated with the Feldman-Cousins method at $95\%$ confidence level~\citep{Feldman1998}. The fit to the proton tail employed to estimate the background contamination is displayed as a black band. Right: The average background contamination for the quoted energy bins in terms of $b$ for a fixed signal efficiency of $50\%$.}
\label{fig:bChoice}
\end{figure}

Because the number of muons on the ground decreases with increasing distance from the shower axis, the observed muon densities exhibit significant statistical uncertainties at large distances due to the finite detector size, independent of the primary particle. Therefore, $M_b$ is defined using the measurements taken with the ``hottest'' station (the one with the highest SD signal) and its six nearest neighbors, which together define the ``hottest'' hexagon of an event. An example of the $M_{b=1}$ distributions for photon- and proton-initiated events is displayed in \cref{fig:bChoice}, left. While there is a clear separation between the two populations, a subset of muon-poor proton events is visible. These events generally correspond to showers with a dominant electromagnetic component, generated by the decay of leading $\pi^0$ produced in first interactions and carrying a substantial fraction of the primary cosmic-ray energy. Vice versa, a minor subset of photon showers with hadronic-like muon content is also observed. These events result from the decay of a leading $\pi^\pm$, caused by a photonuclear interaction, which initiates a hadronic sub-shower early in the shower development.
%Once this limit is reached, the contribution of additional measurements towards discrimination is negligible

%may reach the detector resolution limit, determined by the nominal area,

The choice of $b$ in \cref{eq:Mb} aims to minimize the risk of misclassifying a background event as a photon event, while still ensuring a substantial probability of identifying photon events. The hadronic background is assumed purely of protons to provide a conservative estimation of the photon-hadron separation power, given the existence of a non-negligible contribution of primaries heavier than protons in the cosmic-ray flux~\citep{Auger2024b,Yushkov2019}. We define background contamination as the ratio of the number of proton events below a specified $M_b$ threshold to the total number of proton events. To estimate the former from the available statistics, the $M_b$ of the $10\%$ most photon-like proton events are fitted using an unbinned maximum likelihood procedure, assuming an exponential model (shown by the black line in \cref{fig:bChoice}, left). Then the number of proton events below a given $M_b$ threshold is calculated from the integral of the fitted tail\footnote{This estimation of the background contamination does not depend on the size of the most photon-like background tail, i.e.,\ it remains stable when considering between $5\%$ and $15\%$ of the proton events in the definition of the background tail, thus adding to the robustness of the method.}. The signal efficiency, on the other hand, is defined as the ratio between the photon-initiated events having values of $M_b$ smaller than the threshold value and the total number of photon events.
%the possible muon deficit of the air-shower simulations and the 

As shown in \cref{fig:bChoice}, right, the contamination is minimal for a range of values of $b$ around unity in the energy range of interest for a fixed signal efficiency of $50\%$. It decreases with increasing energy due to the enhanced air-shower muon content. Therefore, $b=1$ is chosen as the index that minimizes the background contamination, defining the discrimination observable as $M_1$. 

% In both cases, all events considered satisfy $\theta < \unit[52]{^\circ}$.
% In both cases, simulated events with $\theta < \unit[52]{^\circ}$ and expected trigger efficiency above $90\%$ have been employed.

\begin{wrapfigure}[17]{I}{0.5\textwidth}
\centering
\includegraphics[width=0.49\textwidth]{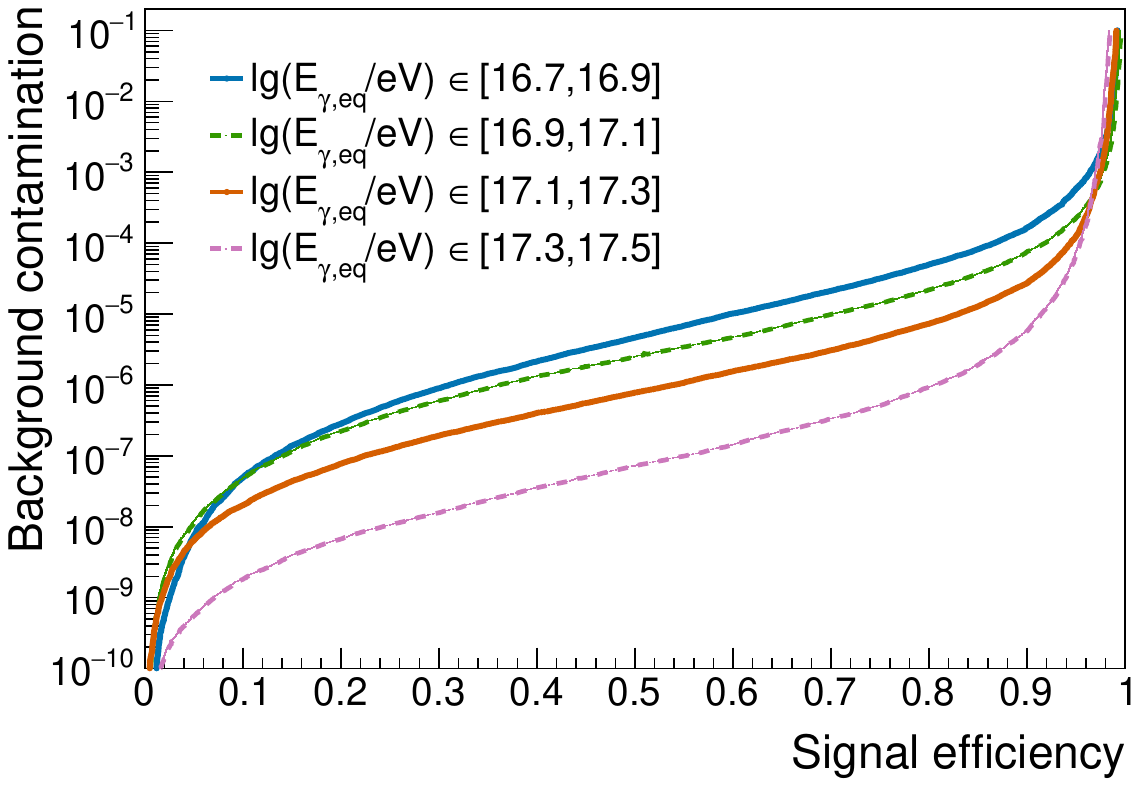}
%\subfigure[]{\includegraphics[width=0.49\textwidth]{images/BCSE_0_52_v2}}	
%\subfigure[\label{fig:candidateCut}]{\includegraphics[width=0.49\textwidth]{images/M1cutE}}
\caption{The background contamination in terms of the signal efficiency for the quoted energy bins in the photon-equivalent scale.}
\label{fig:BCSE}
\end{wrapfigure}

The background contamination and signal efficiency can be computed by scanning over all possible values of the $M_1$ threshold, as exemplified in \cref{fig:BCSE}. The contamination decreases with increasing primary energy, reflecting the increasing number of muons produced in hadronic air showers at higher energies. In addition, a larger signal efficiency can be reached at the expense of a larger background contamination.
%, where the energy assignment for each event is performed using the photon-equivalent scale described in \cref{sec:PES}.

%\footnote{However, this decrease may not always be monotonic. The muon-poor tail of the background distribution is prone to outliers that may occur, especially at large zenith angles where a stronger attenuation of the muon component is encountered. Thus similar background contamination levels across neighboring energy bins may be observed.}

%\begin{figure}

We note that the background contamination and signal efficiency shown in \cref{fig:BCSE} are obtained assuming a complete UMD hexagon, e.g.,\ the north-west hexagon in \cref{fig:UMD433}, left. We verified that the choice of $b=1$ remains effective even when considering missing UMD stations in the first ring of neighbors. However, the discrimination power, i.e. both the background contamination and signal efficiency, varies depending on the real configurations of the detector, as discussed in the next section. 
%using simulated events and

%\section{Photon search in data}
\section{Selection of photon events in data}
\label{sec:application}

%The development of photon-initiated showers is governed predominantly by electromagnetic processes, irrespective of the primary energy. As a result,
%The muon component in photon showers has a slower growth with increasing energy compared to those initiated by hadronic primaries. Therefore, the separation power between photon- and hadron-initiated events improves with increasing primary energy.

In this section, we study the application of the discrimination method to data, considering the actual configuration of the detector. Given that the fraction of photon primaries expected to be present in the cosmic-ray flux may be well below $10^{-5}$, the standard event reconstruction, fine-tuned to hadronic events, is employed for the data. The event selection follows the procedure described in \cref{sec:detectorsAndData}. Both the estimated shower size and zenith angle are used to assign photon-equivalent energy to each event.
%In this way, simulations and data can be compared using the same energy scale. 

The observable $M_1$ quantifies the muon content of each event, as defined in \cref{sec:optimizationMb}, and scales with the number of available UMD stations in the hottest hexagon. Air showers with low muon content, charactistic of photon-like events, would have smaller values of $M_1$ compared to hadronic events. However, if any UMD stations are missing, either due to not being deployed yet at the time of detection or to technical issues during data acquisition, less information on the muon content would be available. This results in smaller values of $M_1$ when fewer than seven UMD stations are operational, leading to an increase of the likelihood of misidentifying background events as photon-initiated. Thus, before applying the discrimination method to the selected data set, it is important to evaluate the impact of missing UMD modules across the seven stations of a hexagon.

%Particularly, an event with a value of $M_1$ below a certain threshold is flagged as a photon candidate.
%General features of ROC curves. Photon candidate cut. Introduce the six categories of events. Bkg contamination parametrization. Burn analysis
%\subsection{Expected discrimination power in data}

Due to the irregular placement of UMD stations in the SD-433, as seen in \cref{fig:UMD433}, left, the muon component arriving at the ground is sampled by a variable number of UMD stations. Therefore, the threshold value of $M_1$ to classify an event as a photon candidate, i.e.,\ the photon candidate cut $M_1^\text{cut}$, is obtained according to the possible configurations in the hottest hexagon. The events are classified in six categories, as listed in \cref{tab:eventCategories}, depending on the following three criteria: the number of available UMD stations (i.e.,\ stations with at least one active module), the total area spanned by these stations, and the relative position of any missing stations (i.e.,\ stations without active modules). The hottest station must contain three active UMD modules regardless of the event category.

%, making the background distribution more similar to the photon distribution
%Given the layout of the UMD stations, events from categories I and II can only be acquired in the north-west hexagon. Thus the minimum sensitive area in the first ring of stations for these two categories is slightly larger than $6 \times \unit[30]{m^2}$ or $5 \times \unit[30]{m^2}$.
%a minimum UMD coverage in the hottest hexagon is required, and

\begin{table}[!tb]
\begin{center}
\begin{tabular}{c c c c c}
\toprule
{Cat.} & \makecell{No. of first-ring\\UMD stations} & \makecell{Minimum UMD area\\in first ring (m$^2$)} & \makecell{Missing UMD\\stations in first ring} & \makecell{No. of\\events} \\ 
\midrule
\midrule
{I} & {$6$} & {$190$} & {$0$} & {3,295} \\
{II} & {$5$} & {$140$} & {$1$} & {1,491} \\
{III} & {$4$} & {$110$} & {$2$ non-NN} & {8,417} \\
{IV} & {$4$} & {$110$} & {$2$ NN} & {298} \\
{V} & {$3$} & {$80$} & {$3$ non-NN} & {1,016} \\
{VI} & {$3$} & {$80$} & {$2$ NN + $1$ non-NN} & {1,402} \\
\midrule
{Total} & {} & {} & {} & {15,919} \\
\bottomrule
\end{tabular}
\end{center}
\caption{\label{tab:eventCategories} The six categories of events based on the available UMD stations in the hottest hexagon, the total detector area, and the relative location of missing UMD stations in the first ring, either nearest neighbors (NN) or non nearest neighbors (non-NN). The last column contains the size of each data subset.}
\end{table}
%, with the size of the burnt sample between parentheses (see text for details)

\begin{figure}
\includegraphics[width=0.49\textwidth]{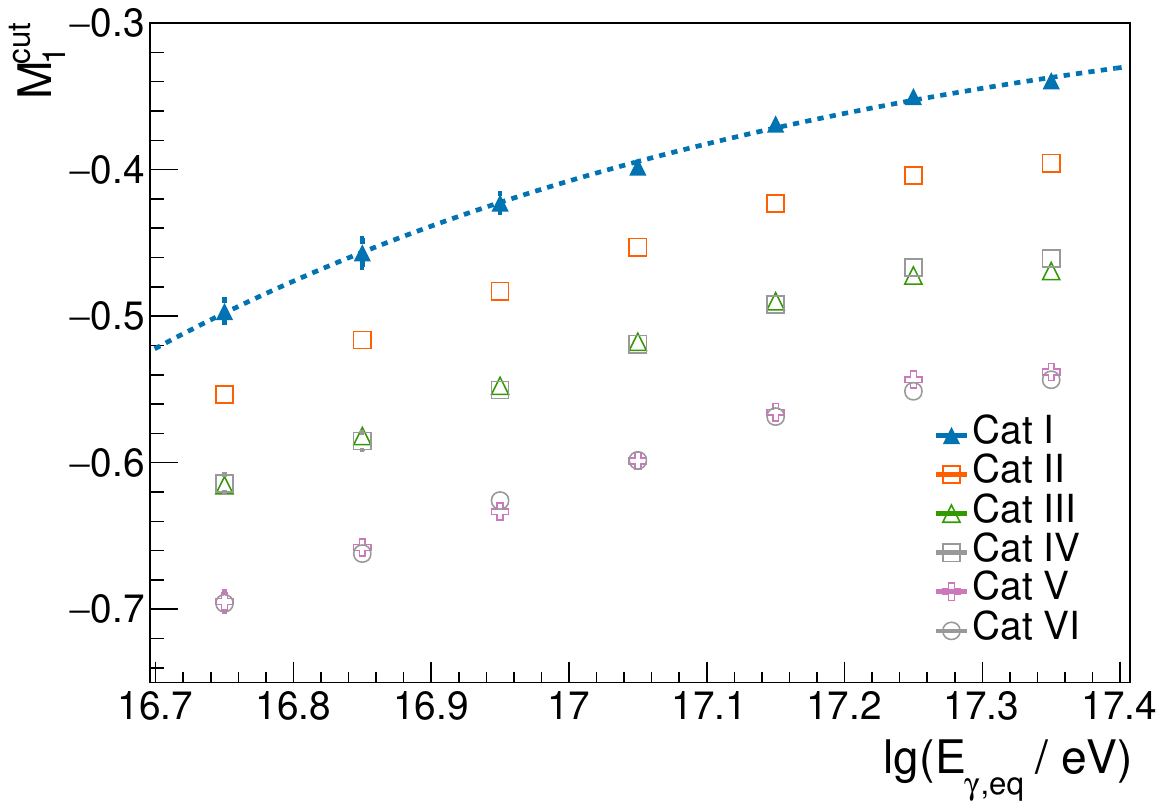}\includegraphics[width=0.49\textwidth]{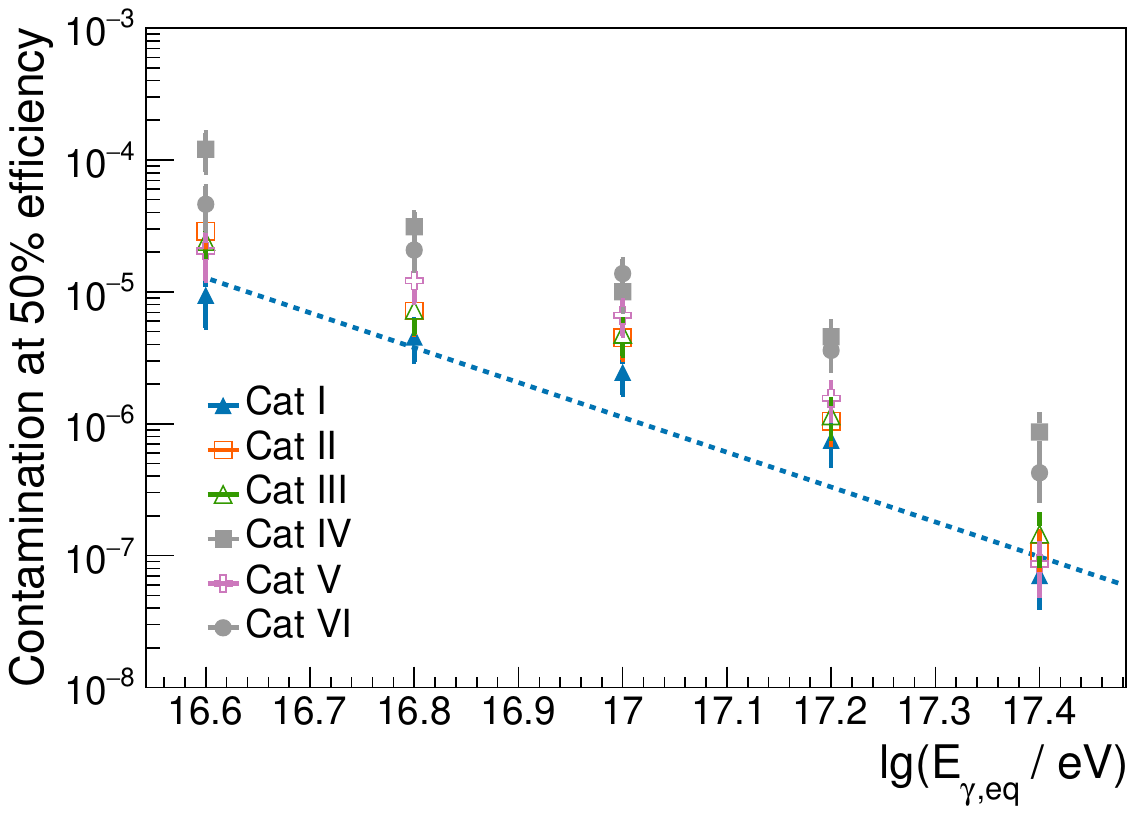}
\caption{The photon candidate cut (left) and the background contamination at $50\%$ signal efficiency (right) in terms of the photon-equivalent energy for each of the event categories cited in \cref{tab:eventCategories}. The dashed lines show examples of the fitted power-law and exponential models of the logarithmic energy, respectively.}
\label{fig:BCCandCutCats}
\end{figure}

The discrimination power of a hexagonal grid that mimics the SD-433 is evaluated through simulations (see \cref{sec:triggerEff}), assuming each SD station is paired with a \unit[30]{m$^2$} UMD station, except for the aforementioned \unit[50]{m$^2$} station. UMD stations or individual modules are randomly masked to simulate missing measurements and assess the discrimination power for each event category. $M_1^\text{cut}$ is defined as the threshold value at which the signal efficiency reaches $50\%$, determined as the median of the $M_1$ distribution for simulated photon events in energy bins. Such signal efficiency is suitable to reach background contamination levels smaller than $10^{-4}$ in all event categories. Therefore, an event is tagged as a photon candidate if $M_1 < M_1^\text{cut}$. As shown in \cref{fig:BCCandCutCats}, left, $M_1^\text{cut}$ increases with primary energy and with the number of UMD stations in the hottest hexagon . The evolution of $M_1^\text{cut}$ is described by a function of the logarithmic energy for each event category and reported in \cref{sec:BC_and_CandCuts}. As depicted in \cref{fig:BCCandCutCats}, right, the contamination remains below $10^{-5}$ for events in category I, while it increases as the number of first-ring UMD stations decreases, with the relative position of missing UMD stations having a significant impact. For instance, the contamination in events with two missing neighboring stations is two to four times larger than in events with non-neighboring missing stations. The background contamination is modeled using an exponential function of logarithmic energy, as detailed in \cref{sec:BC_and_CandCuts}.

%The total area spanned by UMD modules in the first ring has a negligible effect on the background contamination within the same event category.
%Given the correlation of these two quantities, as displayed in \cref{fig:BCSE}, the signal efficiency is typically fixed at a desired level, and the background contamination, which is a suitable metric for evaluating the photon-hadron separation power, is analyzed as a dependent variable.
%, reflecting the slight increase in the muon content of photonic showers
%, and it is independent of the relative position of missing stations (e.g.,\ the values of $M_1^\text{cut}$ for categories III and IV are compatible)

% (b) The photon candidate cut, determined as the median of the $M_1$ distribution for simulated photon events, in terms of the photon-equivalent energy. The solid line represents a power-law function fitted to the binned simulated samples.

%Expected contamination vs E for each event category.
The parameterization of the background contamination for each event category can be employed to obtain a conservative estimation of the number of hadronic events in data that would be misidentified as photon candidates. For instance, the expected number of fake photon events in the category I subset can be computed as $(4.04 \pm 0.04)\times 10^{-3}$ by integrating the blue curve in \cref{fig:BCCandCutCats}, right, above \unit[$10^{16.7}$]{eV}~\citep{GonzalezICRC2023}. Since 3,295 events above \unit[$10^{16.7}$]{eV} fall in that event category, this is equivalent to one fake photon event approximately every 815,650 events\footnote{The expected number of fake photon events in the category IV and VI subsets, corresponding to the configurations with the largest background contamination, is $3.12\times10^{-3}$ and $8.20\times10^{-3}$ respectively. It can be translated to one fake photon event every 95,460 and 170,890 events in each of the subsets.}. Considering that the category I events were acquired over $\unit[\sim10]{months}$ of the observation time, this suggests that one fake photon event would be expected every \unit[106]{years} per hexagon.

%This contamination can be considered a conservative estimate of the probability that a hadronic event in the selected data set is misidentified as a photon candidate.
%Given the number of events corresponding to an event category within the selected data set, the number of background events incorrectly labeled as photon candidates in terms of the photon-equivalent energy can be estimated.

\begin{figure}
%\begin{wrapfigure}[11]{O}{0.5\textwidth}
\centering
\includegraphics[width=0.7\textwidth]{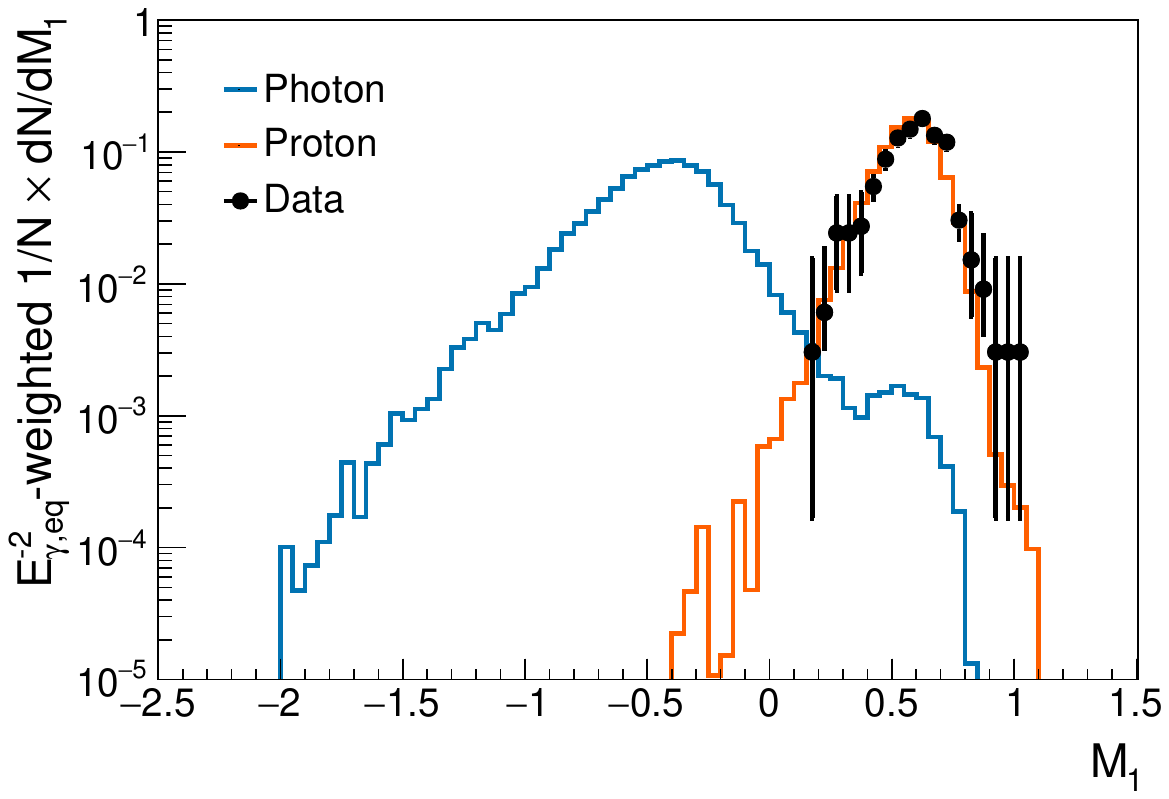}
\caption{The $M_1$ distributions for simulated proton (orange), and photon (blue) events weighted by $E_{\gamma,\text{eq}}^{-2}$ for energies above $\unit[10^{16.7}]{eV}$. Black markers represent the $M_1$ distribution for the events belonging to the burnt data corresponding to the category I subset. The uncertainty in bins with fewer than ten entries is calculated with the Feldman-Cousins method at $95\%$ confidence level~\citep{Feldman1998}.}
\label{fig:BurntM1}
%\end{wrapfigure}
\end{figure}

%[Burn data] Show compatibility of M1 distribution of burn data with the simulated events.
Given the low probability of observing a background event consistent with the photon hypothesis, a fraction of the selected data, referred to as the burnt data sample, is unblinded to check that the $M_1$ distribution in data agrees with simulations, before carrying on the unblinding of the full selected data set. The burnt data represent a fraction $f_\text{burnt}=0.1$ of the selected data set. The $M_1$ distributions of burnt data for each event category are compared with the simulated distributions, as exemplified in \cref{fig:BurntM1}. Simulated events are weighted by $E_{\gamma,\text{eq}}^{-2}$ to obtain energy-integrated distributions, while no weighting is applied to the burnt data sample. As expected, the $M_1$ distribution for the burnt data is compatible with a hadronic origin. It is important to note that the photon-equivalent energy assigned to a data event is around $\sim30\%$ higher than that assigned using the hadronic-optimized energy scale. This discrepancy arises from the larger muon content in cosmic-ray showers compared to the photon showers used to define the photon-equivalent energy calibration. As a result, the reference muon density $\rho_\text{pr}$ used in the $M_1$ definition (see \cref{eq:Mb}) is shifted to higher values in data compared to simulations, leading to smaller $M_1$ values in the former case. Thus, the distribution for data is artificially shifted towards the simulated proton distribution. Nevertheless, the $M_1$ observable is optimized to discriminate between events with significantly different muonic components -such as photons versus all hadronic species- due to the logarithm scaling in its definition, making it largely insensitive to the primary hadronic mass. This limitation does not impact the search for photon events in data, as discussed in the next section.

%, as indicated by the numbers between parentheses in the last column of \cref{tab:eventCategories}
%Thus, distributions of primaries heavier than protons in simulations and data partially overlap with the proton distribution.

%\subsection{Application of the method to the data}
\section{Results of the photon search}
\label{sec:results}
%Unblinding. Show M1 vs E and the photon cut for some categories

\begin{figure}
\centering
\includegraphics[width=0.325\textwidth]{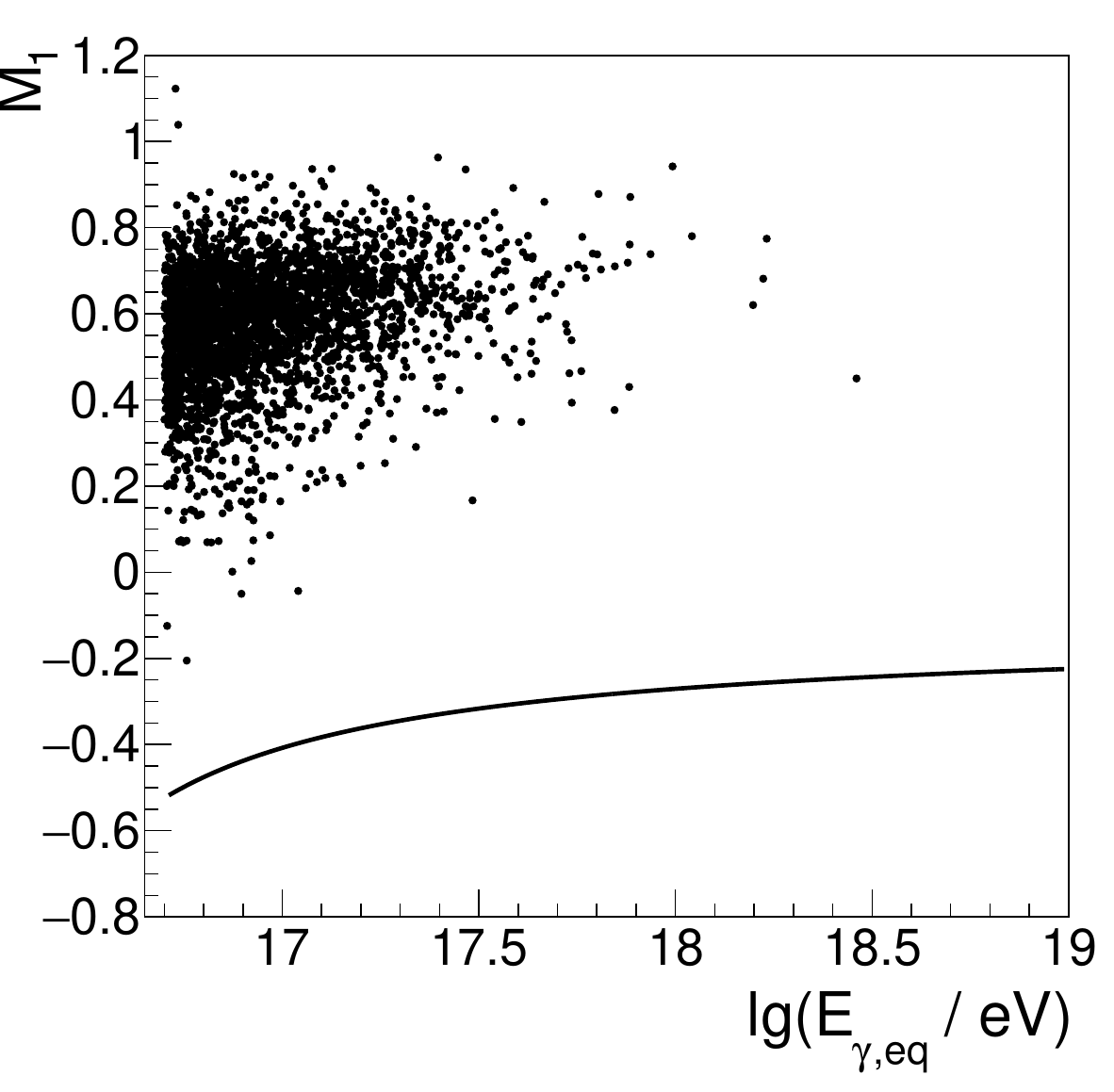} \includegraphics[width=0.325\textwidth]{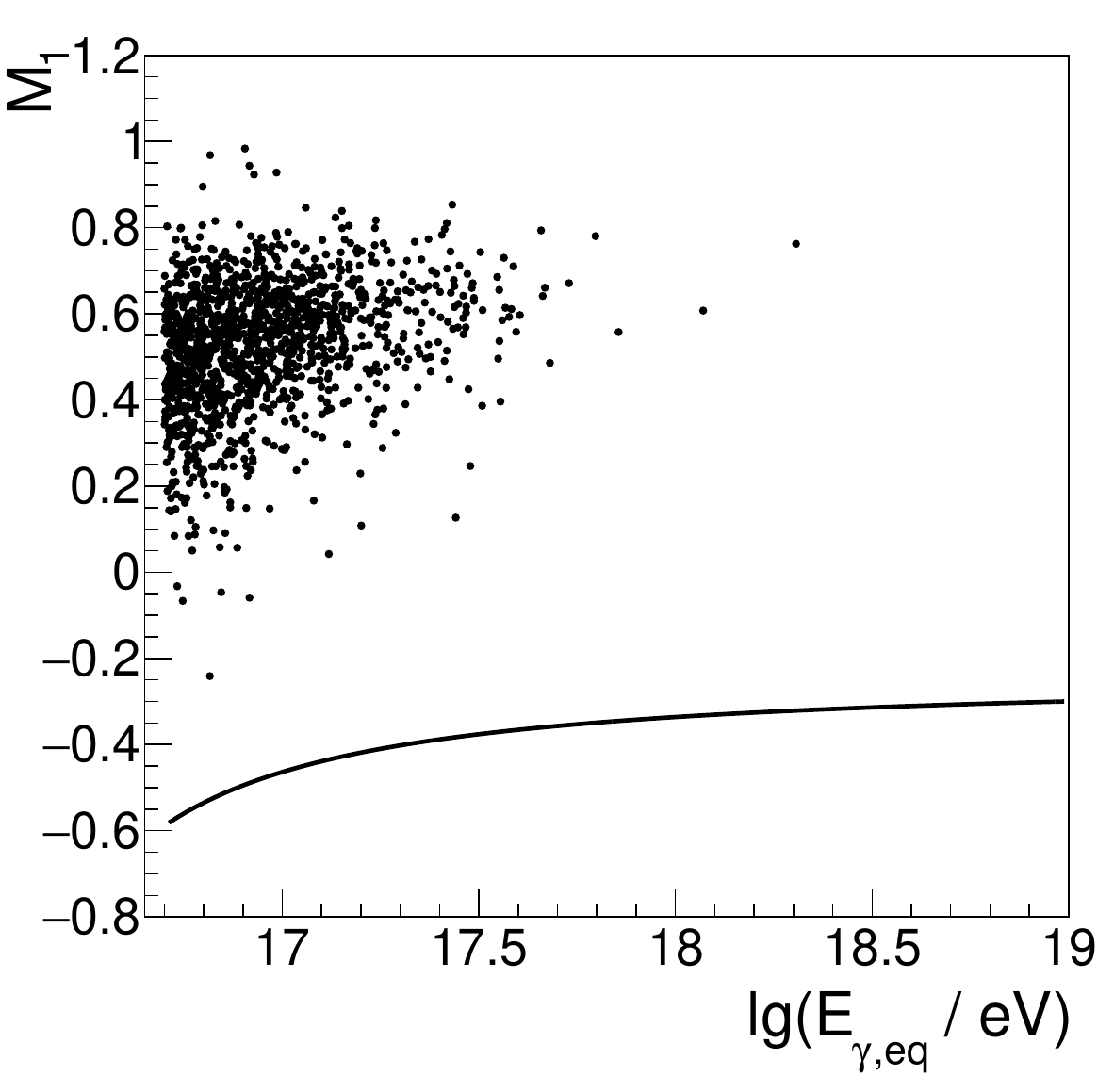}
\includegraphics[width=0.325\textwidth]{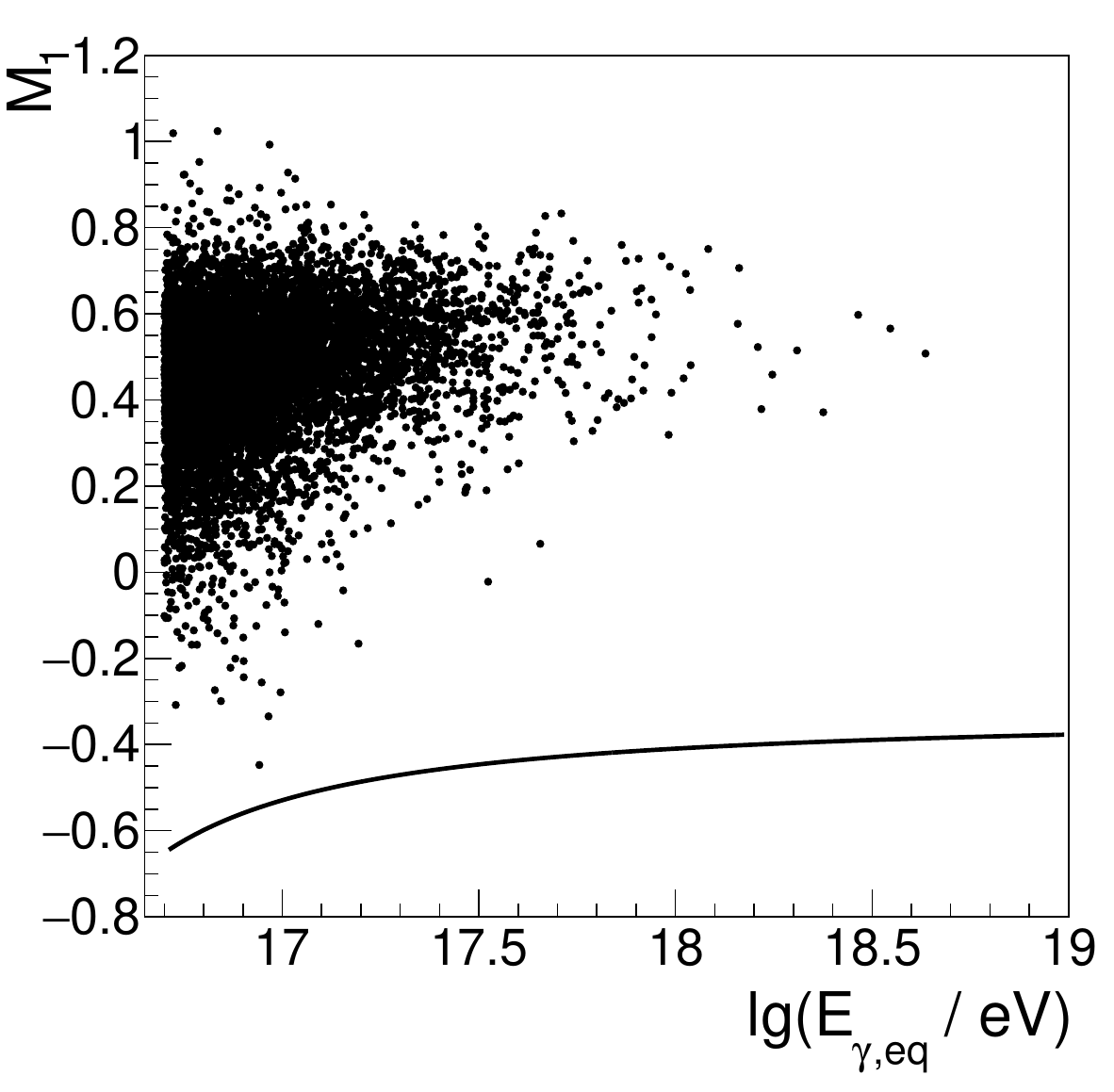}
\\
\includegraphics[width=0.325\textwidth]{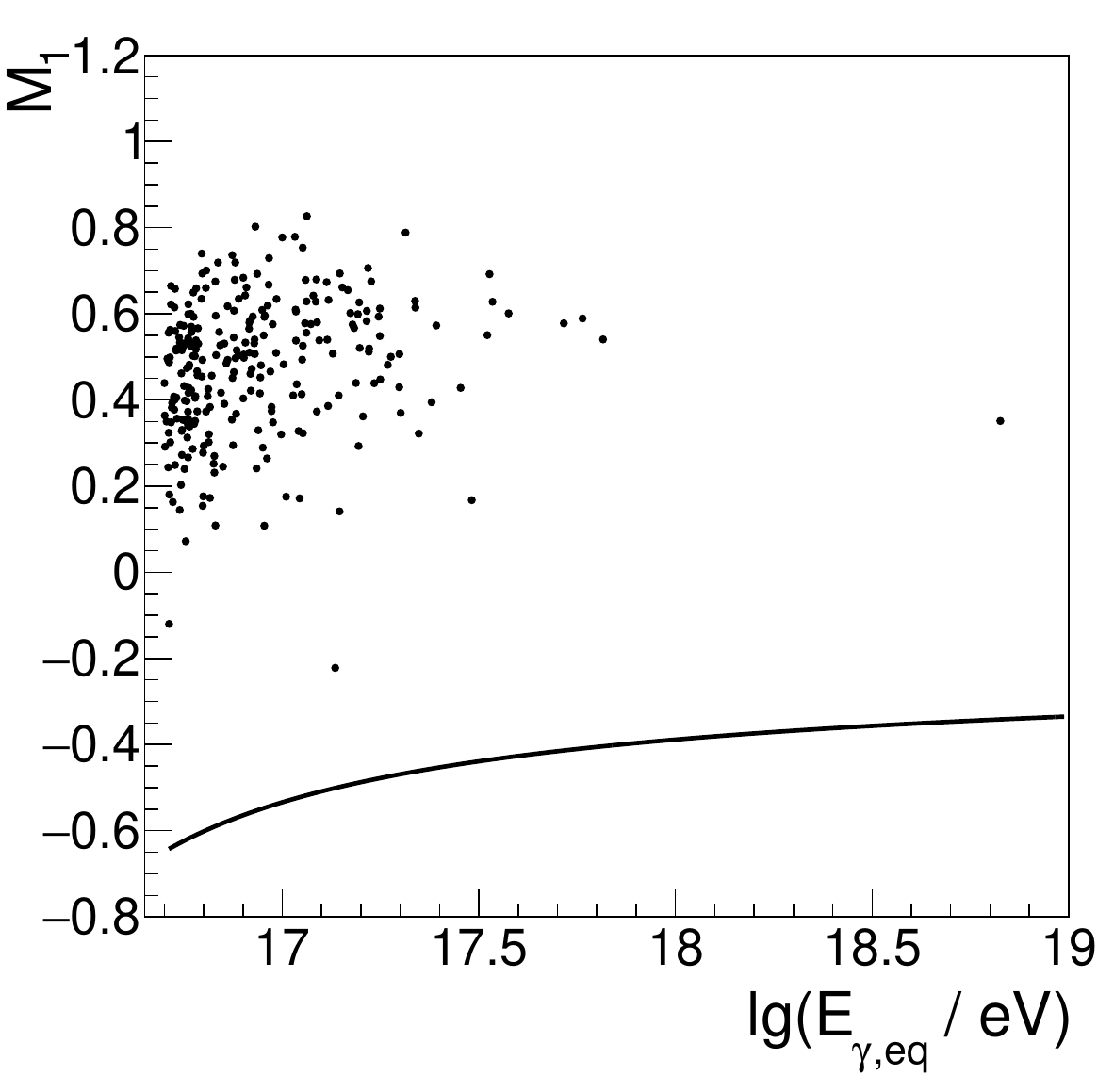} \includegraphics[width=0.325\textwidth]{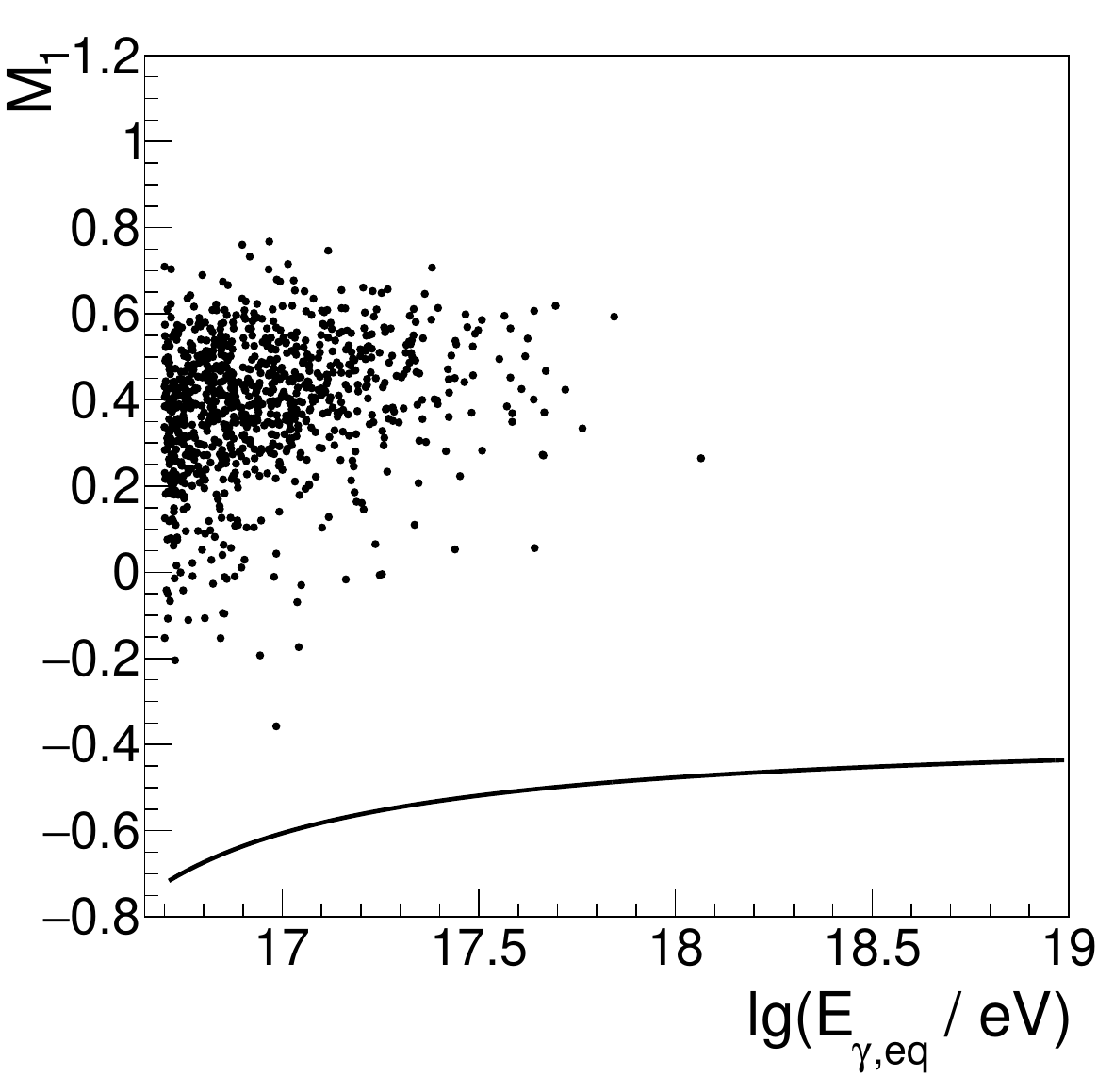} \includegraphics[width=0.325\textwidth]{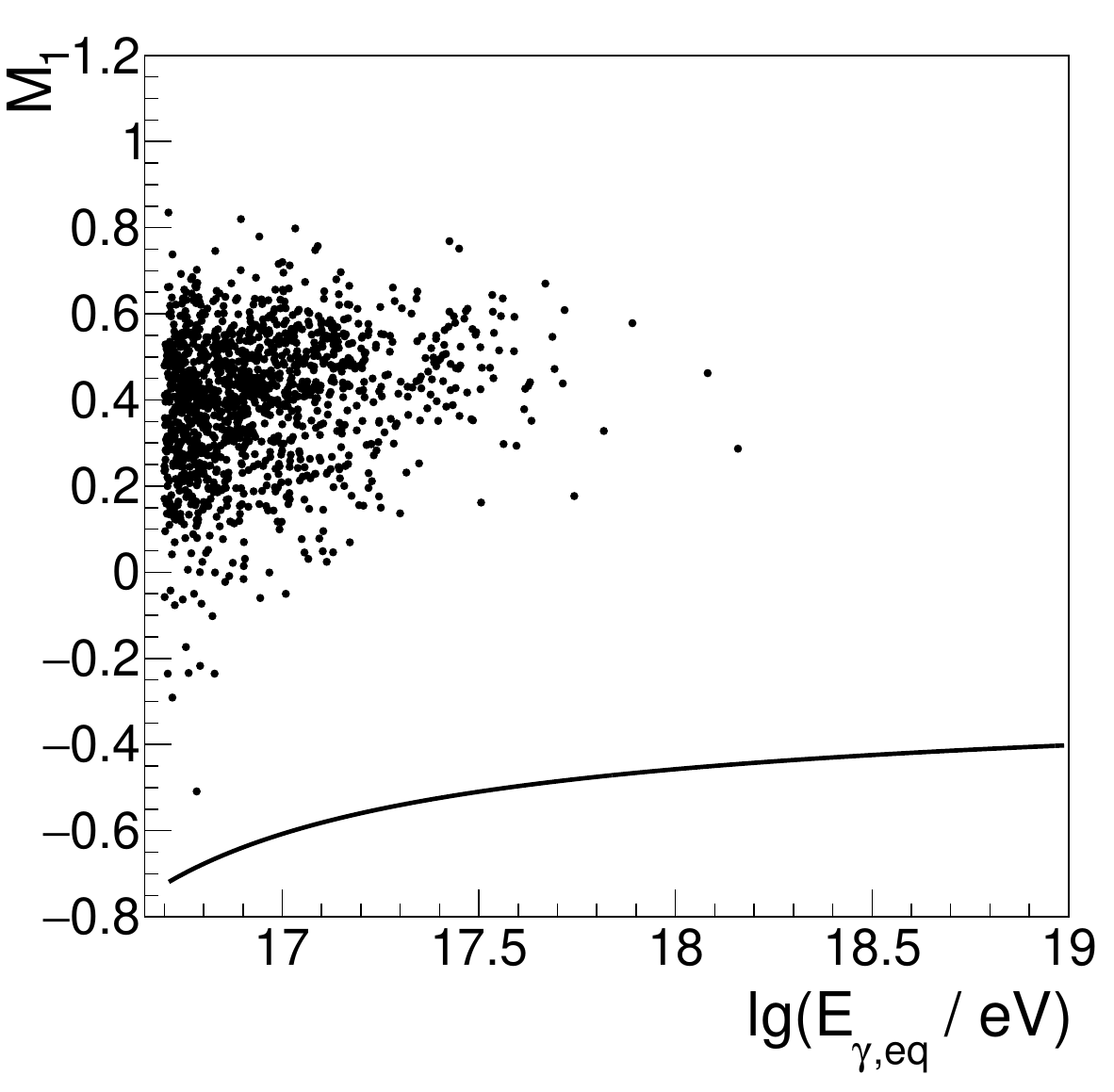}
\caption{The observable $M_1$ in terms of the photon-equivalent energy for events in the search data set. Each panel corresponds to the quoted categories as listed in \cref{tab:eventCategories} from left to right, and top to bottom. Solid lines represent the parametrized photon candidate cut for each category.}
\label{fig:unblindedData}
\end{figure}

Excluding the burnt sample, the remaining $14,299$ events constitute the search data set. For each event, the $M_1$ observable is compared to the parametrized photon candidate cut $M_1^\text{cut}$ for its respective event category, as shown in \cref{fig:unblindedData}. The parameterizations of $M_1^\text{cut}$, performed with simulated events up to \unit[$10^{17.5}$]{eV}, have been extrapolated to higher energies for comparison with the data. Nevertheless, all events exhibit $M_1$ values well above the parametrized $M_1^\text{cut}$, indicating that no photon candidate events are identified.

\subsection{Upper limits calculation}

From the absence of photon flux measurements, upper limits on the integral photon flux are calculated. These limits are inversely proportional to the exposure of the detection system, which accounts for both the operational uptime of the detectors and the efficiency in detecting photon events.

%\begin{figure}
%\centering
%\subfigure[\label{fig:UnblindedCatI}Cat I]{\includegraphics[width=0.49\textwidth]{images/M1E_unblinded_CatI_v2.pdf}}
%\subfigure[\label{fig:UnblindedCatII} Cat II]{\includegraphics[width=0.49\textwidth]{images/M1E_unblinded_CatII_v2.pdf}}
%\subfigure[\label{fig:UnblindedCatIII} Cat III]{\includegraphics[width=0.49\textwidth]{images/M1E_unblinded_CatIII_v2.pdf}}
%\subfigure[\label{fig:UnblindedCatIV} Cat IV]{\includegraphics[width=0.49\textwidth]{images/M1E_unblinded_CatIV.pdf}}
%\subfigure[\label{fig:UnblindedCatV} Cat V]{\includegraphics[width=0.49\textwidth]{images/M1E_unblinded_CatV.pdf}}
%\subfigure[\label{fig:UnblindedCatVI} Cat VI]{\includegraphics[width=0.49\textwidth]{images/M1E_unblinded_CatVI.pdf}}
%\caption{The observable $M_1$ in terms of the photon-equivalent energy for events in the search data set. Each panel corresponds to the quoted categories as listed in \cref{tab:eventCategories}. Solid lines represent the parametrized photon candidate cut for each event subset.}
%\label{fig:unblindedData}
%\end{figure}

%Exposure. Hexagon criterium. Exposure using 5T5+6T5 events and trigger efficiency correction.
The accumulated exposure to photon primaries is derived from the continuous, real-time monitoring of the SD-433 array, as described in \cref{sec:detectorsAndData}. The operational status of each SD station is known every second, allowing stations to be classified as either ``on'' or ``off''. This enables real-time monitoring of the status of each hexagon\footnote{Technical issues affecting eight out of the $19$ SD stations during $16$ distinctive periods (lasting between one day and one week) were identified, accounting for less than $1\%$ of the operation time. During these periods, the stations were considered ``off''.}. Following the 5T5 and 6T5 event selection criteria, a hexagon is considered active if at least five ``on'' WCDs are surrounding the hottest one. When a 6T5 condition is met, the effective area of a hexagon with a side length $d=\unit[433]{m}$ is the one of the unitary hexagon, i.e., $A=\sqrt{3}/2\,\times\,d^2$. For the 5T5 condition, the area is reduced by a geometrical factor of $1/3$ to account for the missing nearest-neighbor station. Consequently, an active hexagon contributes to the exposure calculation with a time-dependent area, $A(t)$, reflecting its real-time status. 
%\footnote{As explained in \cref{sec:detectorsAndData}, all the WCDs and UMD modules are manually flagged as ``off'' during known periods of unstable operations.}

As discussed in \cref{sec:application}, a minimum area spanned by UMD modules is required to adequately sample the air-shower muon content. The number of ``on'' UMD modules at the central station and the first ring around it, as well as the relative position of any missing UMD stations in the first ring, are dynamic. They vary over time due to the absence of deployed modules or technical issues affecting normal operation. These requirements are mapped by a step function, $u(t)$, incorporated into the exposure calculation.

% arriving on ground and achieve sufficient discrimination power
% requirements are the following: i) the central station must have all three modules ``on''; ii) an area of at least \unit[80]{m$^2$} must be spanned by ``on'' modules across the nearest-neighbor stations around the central one; and iii) there are at most two neighboring missing UMD stations in the first ring.

The trigger efficiency for photon-initiated showers, $\tau$, must also be included in the exposure calculation. This efficiency depends on the primary energy and zenith angle, as discussed in \cref{sec:photonCal}, and the instantaneous exposure decreases as the primary energy decreases. 
%to account for the probability of generating a T3 event trigger for photon primaries

The accumulated exposure to photon primaries is determined by integrating the effective area of each hexagon over the entire observation period, $t$, above a given photon-equivalent energy, $E_\gamma^\text{th}$, and over the solid angle, $\Omega$, as:

\begin{equation}
\epsilon_\gamma = \sum_{\text{hexagons}} \int_{E_\gamma^\text{th}} \int_{\Omega} \int_{t} A(t) \cos\theta \times u(t) \times \tau(E_{\gamma,\text{eq}},\theta) \, \mathrm{d}t \, \mathrm{d}\Omega \, \mathrm{d}E_{\gamma,\text{eq}}.
\label{eq:expo}
\end{equation} 

\noindent The trigger efficiency can be integrated with a weight proportional to a $E_\gamma^{-2}$ spectrum, motivated by theoretical predictions of the diffuse galactic photon flux~\citep{Berat2022}, top-down models~\citep{Aloisio2006} and previous photon searches by the Pierre Auger Collaboration~\citep{Auger2022}.
%as computed for the average flux of events per active SD-433 hexagon.
%and corrected a-posteriori to account for the trigger efficiency for photon-initiated events developed in \cref{sec:photonCal}
%The accumulated exposure in the full trigger efficiency regime and up to a zenith angle of \unit[52]{$^\circ$} is $\epsilon_\gamma = \unit[\left(0.63\pm0.03\right)]{km^2\,sr\,yr}$.
% It is characterized by the number of on UMD stations in the first crown around the SD station with the maximum signal (see \cref{sec:dataSelection}), the total area spanned by their on modules, and the relative position of the missing ones. In addition, the selected events must contain the three operational UMD modules around the SD station with the maximum signal.

The upper limit to the integral photon flux, $\Phi_\gamma$, is calculated as:

\begin{equation}
\label{eq:upperLimits}
\Phi_\gamma < \frac{3.095}{\left(1-f_\text{burnt}\right) \times f_\text{cut,$\gamma$} \times \epsilon_{\gamma}}
\end{equation}

%Calculation of the upper limits. 
\noindent where the numerator represents the upper limit to the number of photon-compatible events calculated using the Feldman-Cousins approach~\citep{Feldman1998} given the null observation and a confidence level of $95\%$. The exposure to photon primaries above given energy, $\epsilon_{\gamma}$, is reduced by a factor $1-f_\text{burnt}$ by construction and a factor $f_\text{cut,$\gamma$}=0.504$, reflecting the average signal efficiency of the $M_1^\text{cut}$ parameterizations across all event categories. Above threshold energies of \mbox{$50$, $80$, $120$ and \unit[200]{PeV}}, the integrated exposure up to a zenith angle of \unit[52]{$^\circ$} is $\epsilon_\gamma = (0.58\pm0.02)$, $(0.61\pm0.03)$, $(0.63\pm0.03)$ and \unit[$(0.63\pm0.03)$]{km$^2$\,sr\,yr}, respectively, in which the dependence on the threshold energy comes in from the photon trigger efficiency model\footnote{The model for the trigger efficiency was computed in terms of $E_\text{MC}$. Thus, the energy bias between $E_{\gamma,\text{eq}}$ and $E_\text{MC}$, as explained in \cref{sec:PES}, was introduced before performing the integration.}. An uncertainty of $4\%$ is assigned to the exposure rising from the uncertainty of the event rate per active SD-433 hexagon. The upper limits are computed under a conservative scenario, where the average exposure above each threshold energy is decreased by its uncertainty, resulting in integral upper limits of $12.3$, $11.7$, $11.3$ and \unit[$11.3$]{km$^{-2}$\,sr$^{-1}$\,yr$^{-1}$} above the mentioned threshold energies.

%, related to the burn data set size,

\begin{figure}
\centering
\includegraphics[width=0.8\textwidth]{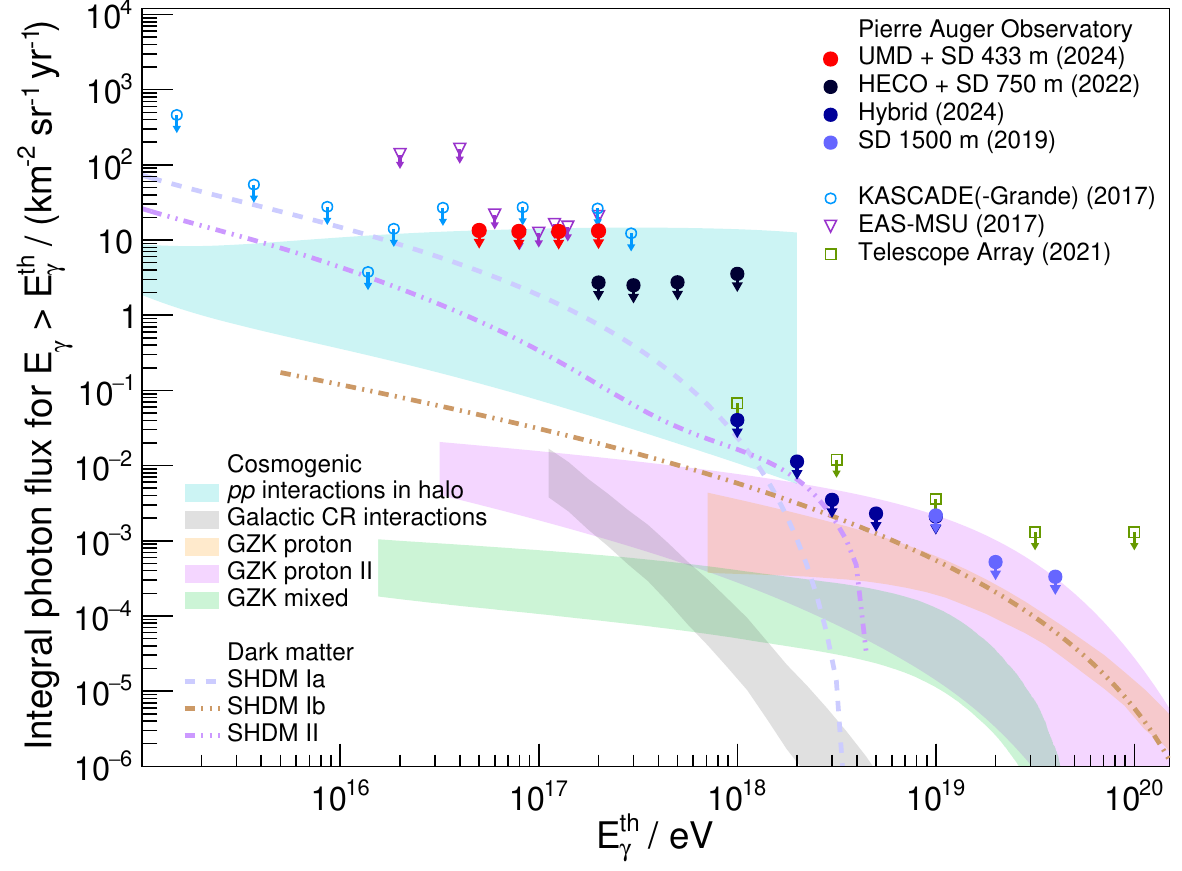}
\caption{The upper limits to the integral photon flux above a threshold energy $E_{\gamma}^\text{th}$ obtained in this work at a $95\%$ confidence level, as solid red markers, and limits obtained by the Pierre Auger Collaboration at higher energies with a $95\%$ confidence level~\citep{Niechciol2023,Auger2024g}, as solid blue and black markers, as well as limits reported by other experiments established at $90\%$ confidence level, except those obtained by Telescope Array at $95\%$ confidence level (see text for a full list of references). The light-colored bands represent predictions of cosmogenic fluxes: from interactions between UHE cosmic rays and the interstellar galactic matter~\citep{Berat2022} (in gray), with background radiation fields~\citep{Sarkar2011,Gelmini2022,Bobrikova2021} (in violet, green, and orange, depending on the quoted primary composition), and with hot gas in the galactic halo~\citep{Kalashev2014} (in blue). Dashed lines correspond to super-heavy dark matter predictions (see text for details).}
\label{fig:limits}
\end{figure}

In \cref{fig:limits}, we present the upper limits on the integral photon flux obtained in this study (represented by red markers), alongside the limits previously reported by the Auger Collaboration~\citep{Niechciol2023,Auger2024g}, by KASCADE-Grande~\citep{KASCADEGrande2017}, by EAS-MSU~\citep{Formin2017}, and by Telescope Array~\citep{TA2019,TA2021}. The limits derived in this work are the only ones based on measurements from the Southern Hemisphere in the tens of PeV energy domain. This analysis extends the photon search program of Auger, covering over three decades of cosmic-ray energy. It is the first one in which direct measurements of air-shower muons provided by the UMD are employed.

%are the most stringent in the energy range between $50$ and \unit[200]{PeV}, calculated at a $95\%$ confidence level, 

Diffuse photon fluxes are expected from the interaction of UHE cosmic rays with background radiation fields~\citep{Sarkar2011,Bobrikova2021,Gelmini2022} and with the interstellar Galactic matter~\citep{Auger2023k}, as discussed in \cref{sec:intro}. These are indicated by the shaded bands in \cref{fig:limits}. The upper limits derived in this study are between two and three orders above these cosmogenic fluxes. However, the expected diffuse gamma-ray flux from proton-proton interactions in the galactic halo~\citep{Kalashev2014} is shown to be within the reach of the upper limits obtained in this work. Moreover, predictions from various phenomena beyond the Standard Model may emerge above these cosmogenic fluxes, e.g.,\ those arising from the decay of super-heavy dark matter particles~\citep{Anchordoqui2021}. In the above-mentioned figure, we show the predicted diffuse fluxes in several scenarios: assuming decay channels into hadrons, a mass $M_\text{X}=\unit[10^{10}]{GeV}$ and a lifetime $\tau_\text{X}=\unit[3\times10^{21}]{yr}$~\citep{Kalashev2016}; $M_\text{X}=\unit[10^{12}]{GeV}$ and $\tau_\text{X}=\unit[10^{23}]{yr}$~\citep{Kalashev2016}; and assuming decay into leptons, a mass $M_\text{X}=\unit[10^{10}]{GeV}$ and a lifetime $\tau_\text{X}=\unit[3\times10^{21}]{yr}$~\citep{Kachelriess2018}. Further sensitivity provided by the next \unit[10]{years} of data taking will allow us to constrain the parameter space of the discussed models thanks to an expected improvement in the upper limits by a factor larger than $20$.

% through the GZK effect
%Another significant source of cosmogenic flux are the interactions of UHE cosmic rays with interstellar Galactic matter~\citep{Berat2022}, represented by the blue-shaded band, particularly below \unit[$10^{17}$]{eV}.
%However, the upper limits derived in this study are nearly three orders of magnitude above the expected GZK flux for a mixed composition consistent with the latest results by the Auger Collaboration.  Our upper limits remain nearly two orders of magnitude above these expectations. 

%Although the density of gas in the Galactic halo is much lower than in the disk, it is sufficient to contribute to the diffuse photon flux through proton-proton interactions. Assuming that the observed IceCube astrophysical neutrino flux originates entirely from the charged pions produced in such interactions, the associated photon flux has been estimated for various source distributions and spectral shape of the neutrino flux~\citep{Kalashev2014}. However, this component is significantly constrained by the current observational photon limits above \unit[$10^{17}$]{eV}~\citep{Auger2022c}.

\subsection{Systematic uncertainties on the upper limits}

Variations in the upper limits can arise from the energy bias related to the photon-equivalent energy scale, the exposure estimation and the changes in the number of photon candidate events.

%The integral photon flux $J$ above threshold energy $E_\gamma^\text{th}$ in the photon-equivalent scale, assuming a power-law spectrum with index $\gamma > 1$ %is given by:

%\begin{equation}
%\label{eq:integralFlux}
%J\left(E_\gamma^\text{th}\right) = c \int_{E_\gamma^\text{th}}^\infty E^{-\gamma}\,dE = - c \times \frac{{E_\gamma^\text{th}}^{1-\gamma}}{1-\gamma}
%\end{equation}

The integral photon flux is a power law of the threshold energy $E_\gamma^\text{th}$, assuming an index $\gamma>1$. If the latter is adjusted to  $E_\gamma^\text{th}\times(1+b(E_\gamma^\text{th}))$ to account for a bias on the energy, $b$, the relative effect on the integral flux, and hence on the upper limits, is:

\begin{equation}
\label{eq:sysEffectBias}
\frac{\Delta \Phi_\gamma}{\Phi_\gamma} = (1+b(E_\gamma^\text{th}))^{1-\gamma}-1
\end{equation}

\noindent Proton events are assigned an overestimated energy in the photon-equivalent scale, with an energy bias that decreases from $10\%$ to $5\%$ in the energy range of interest (see \cref{fig:photonScaleAndPES}, right). Following \cref{eq:sysEffectBias}, the effect of this bias on the upper limits ranges from $-10\%$ to $-6\%$ with increasing energy. Since this would imply a decrease in the upper limits, we neglect this effect to maintain a conservative calculation. Similarly, a negative energy bias between $-10\%$ and $-15\%$, decreasing linearly with the logarithm of energy, is found for photon events in the photon-equivalent energy scale (see \cref{fig:photonScaleAndPES}, right). This bias leads to a relative increase of the upper limits, ranging from $12\%$ to $17\%$ with increasing energy.

%When using an energy scale deduced from the fluorescence measurements above $\unit[10^{17}]{eV}$~\citep{BrichettoICRC2023} the energy for photon events is overestimated by $33\%$. 

The impact of variations in the spectral index assumed in the energy-integrated exposure calculation has been examined. A variation of $\pm0.5$ in the spectral index results in an average change in the exposure of between $0.7\%$ and $1.7\%$ for the quoted threshold energies. Consequently, the upper limits on the integral flux increase by a factor ranging from $0.8\%$ to $2.2\%$ with increasing energy.

The number of photon candidate events can increase if the values of $M_1$ decrease. In the following, we discuss how varying the parameters in the $M_1$ definition might cause this effect. First, the measured muon density can decrease by up to $5.6\%$ when varying the parameters involved in the muon counting strategy used to convert digital traces to the number of muons in the UMD stations~\citep{DeJesusICRC2023}. Second, the UMD simulation assumes a soil density of \unit[2.38]{g\,cm$^{-3}$}, leading to a fixed soil shielding to muons. The standard deviation of the soil density is \unit[0.05]{g\,cm$^{-3}$}~\citep{WundheilerPhD}, which results in a $2.8\%$ increase in the reference muon density $\rho_\text{pr}$~\citep{AMIGAFAL2019}. Lastly, different high-energy hadronic models predict varying numbers of muons in simulations, which could affect $\rho_\text{pr}$. The only model surpassing the employed one, EPOS-LHC, in terms of the predicted number of muons, is Sibyll2.3d by about $5\%$~\citep{Riehn2020}. Combining the three factors, the overall systematic effect translates into an absolute shift of $-0.06$ across all $M_1$ values. Despite this shift, no event falls below the parametrized photon candidate cuts.

Following the conservative approach mentioned before, the upper limits above each threshold energy can be increased by the corresponding factors arising from the bias in the photon-equivalent energy and the spectral index employed in the exposure calculation. Thus, the final upper limits reported in this work, including a shift towards the worst-case scenario motivated by the considered systematic effects, are $13.8$, $13.5$, $13.3$ and \unit[$13.6$]{km$^{-2}$\,sr$^{-1}$\,yr$^{-1}$} for threshold energies of $50$, $80$, $120$ and \unit[200]{PeV}. A summary of the upper limits before and after accounting for the systematic effects is presented in \cref{tab:upperLimits}. These limits translate into constraints on the photon fraction in the measured cosmic-ray flux~\citep{Novotny2021}. Above the specified energy thresholds, the upper limits on the photon fraction are $0.056\%$, $0.14\%$, $0.35\%$ and $0.97\%$ at a confidence level $95\%$.

\begin{table}[!tb]
\begin{center}
\begin{tabular}{c c c c}
\toprule
{$E_\gamma^\text{th}$ (PeV)} & {$\epsilon_\gamma$ (km$^2$\,sr\,yr)} & {$\Phi_\gamma$ (km$^{-2}$\,sr$^{-1}$\,yr$^{-1}$)} & {$\Phi_\gamma \, + \sigma_\text{sys}$ (km$^{-2}$\,sr$^{-1}$\,yr$^{-1}$)} \\ 
\midrule
\midrule
{$50$} & {$(0.58\pm0.02)$} & {$12.3$} & {$13.8$} \\
{$80$} & {$(0.61\pm0.03)$} & {$11.7$} & {$13.5$} \\
{$120$} & {$(0.63\pm0.03)$} & {$11.3$} & {$13.3$} \\
{$200$} & {$(0.63\pm0.03)$} & {$11.3$} & {$13.6$} \\
\bottomrule
\end{tabular}
\end{center}
\caption{\label{tab:upperLimits} The upper limits on the integral photon flux above each quoted energy threshold before (third column) and after (fourth column) accounting for the discussed systematic effects. The exposure above each energy threshold is obtained from \cref{eq:expo}.}
\end{table}

\section{Conclusions and outlook}
\label{sec:concl}
%Make clear what is new and why is it so important. Enumerate findings in decreasing order of importance.
%Bounce back to the intro. "Do you remember I told you about a gap in the literature? This is my answer to that."
%Discuss what the limits imply regarding the UHE photon models. Assess the limitations of your analysis.

The search for primary photons above \unit[10]{PeV} has so far been conducted by experimental facilities exclusively located in the Northern Hemisphere, thereby presenting a restricted exposure towards the Galactic plane. This article presents the first search for a diffuse flux of primary photons above between $50$ and \unit[200]{PeV} from the Southern Hemisphere. At these energies, primary photons may be expected from sources located not much further than the Galactic center. Leveraging the densest surface array and underground muon detectors of the Pierre Auger Observatory, we analyzed a high-quality data set comprising over 15,000 events above \unit[50]{PeV}. The direct measurements of the air-shower high-energy muon component provided excellent photon-hadron separation power, resulting in a probability of incorrectly identifying a proton as a photon primary smaller than $10^{-5}$ when the probability of observing a photon is set at $50\%$, under the conservative assumption of a simulated pure proton background. Notably, no events consistent with a photon origin were found in our dataset.

Consequently, we established upper limits on the integral photon flux above $50$ to \unit[200]{PeV} ranging from $13.3$ to \unit[$13.8$]{km$^{-2}$\,sr$^{-1}$\,yr$^{-1}$} at a $95\%$ confidence level. An average exposure of \mbox{$\unit[\left(0.63\pm0.03\right)]{km^2\,sr\,yr}$}, equivalent to approximately eight months of ideal operation, conservatively reduced by one standard deviation, was employed for the computation of the upper limits. Furthermore, the reported limits were increased to account for systematic uncertainties based on the assumed spectral index of the diffuse photon flux and bias in the predicted photon energy due to the specific scale employed in this analysis.

%As the exposure accumulated by the SD-433 increases from the ongoing operation of the Observatory, it will also be possible to reach the predictions from  thus constraining 

Thanks to the addition of the surface and muon detectors spaced at \unit[433]{m}, the search for primary photons can presently be conducted with data measured by the Auger Observatory spanning over three decades of cosmic-ray energy. The additional insights provided by the complementary detection techniques of the AugerPrime upgrade~\citep{Berat2022a} will offer greater sensitivity in cosmic-ray mass composition studies. As the Observatory continues taking data, the exposure will progressively increase, leading to more stringent limits or possibly the discovery of the most energetic primary photons. Furthermore, the full deployment of UMD stations in the SD-433 array now enables the use of all seven hexagons in the array for photon searches. The expected exposure until the planned end of operations of the Auger Observatory will provide a significant opportunity to constrain the mass-lifetime phase-space for specific super-heavy dark matter models and to explore the expected photon flux from proton-proton interactions in the Galactic halo at tens of PeV. In addition to the search for a diffuse photon flux, this study lays the groundwork for a nearly real-time search for primary photons in the tens of PeV domain, enhancing the role of the Auger Observatory in the global multi-messenger astrophysics community.

%, particularly in the quest for an ultra-high-energy primary photon signal in the cosmic-ray flux

%This effort holds the potential to enhance the role of the Auger Observatory role in the global multi-messenger astrophysics community. Additionally, this study lays the groundwork for a real-time search for primary photons at the tens of PeV domain using the Auger Observatory. Such a search can be coordinated with alerts from other multi-messenger observatories. 

%The exposure accumulated by Auger until 2035 can be estimated taking into account that in September 2023 the central stations of the three hexagons would provide enough sensitivity (light red line) to put constraints in the mass-lifetime phase-space for specific dark matter models (dashed lines), as well as to explore the expected photon flux from pp interactions in the Galactic halo.

%This study will allow us to constrain further the theoretical dark-matter models, the properties of cosmic-ray sources, and the high-energy regime of the Galactic gamma-ray sources, leveraging the privileged exposure of Auger towards the Galactic plane.

%\clearpage

% <!-----------------!> %

%\section*{Acknowledgments}
% created on 2025-01-15
\section*{Acknowledgments}

\begin{sloppypar}
The successful installation, commissioning, and operation of the Pierre
Auger Observatory would not have been possible without the strong
commitment and effort from the technical and administrative staff in
Malarg\"ue. We are very grateful to the following agencies and
organizations for financial support:
\end{sloppypar}

\begin{sloppypar}
Argentina -- Comisi\'on Nacional de Energ\'\i{}a At\'omica; Agencia Nacional de
Promoci\'on Cient\'\i{}fica y Tecnol\'ogica (ANPCyT); Consejo Nacional de
Investigaciones Cient\'\i{}ficas y T\'ecnicas (CONICET); Gobierno de la
Provincia de Mendoza; Municipalidad de Malarg\"ue; NDM Holdings and Valle
Las Le\~nas; in gratitude for their continuing cooperation over land
access; Australia -- the Australian Research Council; Belgium -- Fonds
de la Recherche Scientifique (FNRS); Research Foundation Flanders (FWO),
Marie Curie Action of the European Union Grant No.~101107047; Brazil --
Conselho Nacional de Desenvolvimento Cient\'\i{}fico e Tecnol\'ogico (CNPq);
Financiadora de Estudos e Projetos (FINEP); Funda\c{c}\~ao de Amparo \`a
Pesquisa do Estado de Rio de Janeiro (FAPERJ); S\~ao Paulo Research
Foundation (FAPESP) Grants No.~2019/10151-2, No.~2010/07359-6 and
No.~1999/05404-3; Minist\'erio da Ci\^encia, Tecnologia, Inova\c{c}\~oes e
Comunica\c{c}\~oes (MCTIC); Czech Republic -- GACR 24-13049S, CAS LQ100102401,
MEYS LM2023032, CZ.02.1.01/0.0/0.0/16{\textunderscore}013/0001402,
CZ.02.1.01/0.0/0.0/18{\textunderscore}046/0016010 and
CZ.02.1.01/0.0/0.0/17{\textunderscore}049/0008422 and CZ.02.01.01/00/22{\textunderscore}008/0004632;
France -- Centre de Calcul IN2P3/CNRS; Centre National de la Recherche
Scientifique (CNRS); Conseil R\'egional Ile-de-France; D\'epartement
Physique Nucl\'eaire et Corpusculaire (PNC-IN2P3/CNRS); D\'epartement
Sciences de l'Univers (SDU-INSU/CNRS); Institut Lagrange de Paris (ILP)
Grant No.~LABEX ANR-10-LABX-63 within the Investissements d'Avenir
Programme Grant No.~ANR-11-IDEX-0004-02; Germany -- Bundesministerium
f\"ur Bildung und Forschung (BMBF); Deutsche Forschungsgemeinschaft (DFG);
Finanzministerium Baden-W\"urttemberg; Helmholtz Alliance for
Astroparticle Physics (HAP); Helmholtz-Gemeinschaft Deutscher
Forschungszentren (HGF); Ministerium f\"ur Kultur und Wissenschaft des
Landes Nordrhein-Westfalen; Ministerium f\"ur Wissenschaft, Forschung und
Kunst des Landes Baden-W\"urttemberg; Italy -- Istituto Nazionale di
Fisica Nucleare (INFN); Istituto Nazionale di Astrofisica (INAF);
Ministero dell'Universit\`a e della Ricerca (MUR); CETEMPS Center of
Excellence; Ministero degli Affari Esteri (MAE), ICSC Centro Nazionale
di Ricerca in High Performance Computing, Big Data and Quantum
Computing, funded by European Union NextGenerationEU, reference code
CN{\textunderscore}00000013; M\'exico -- Consejo Nacional de Ciencia y Tecnolog\'\i{}a
(CONACYT) No.~167733; Universidad Nacional Aut\'onoma de M\'exico (UNAM);
PAPIIT DGAPA-UNAM; The Netherlands -- Ministry of Education, Culture and
Science; Netherlands Organisation for Scientific Research (NWO); Dutch
national e-infrastructure with the support of SURF Cooperative; Poland
-- Ministry of Education and Science, grants No.~DIR/WK/2018/11 and
2022/WK/12; National Science Centre, grants No.~2016/22/M/ST9/00198,
2016/23/B/ST9/01635, 2020/39/B/ST9/01398, and 2022/45/B/ST9/02163;
Portugal -- Portuguese national funds and FEDER funds within Programa
Operacional Factores de Competitividade through Funda\c{c}\~ao para a Ci\^encia
e a Tecnologia (COMPETE); Romania -- Ministry of Research, Innovation
and Digitization, CNCS-UEFISCDI, contract no.~30N/2023 under Romanian
National Core Program LAPLAS VII, grant no.~PN 23 21 01 02 and project
number PN-III-P1-1.1-TE-2021-0924/TE57/2022, within PNCDI III; Slovenia
-- Slovenian Research Agency, grants P1-0031, P1-0385, I0-0033, N1-0111;
Spain -- Ministerio de Ciencia e Innovaci\'on/Agencia Estatal de
Investigaci\'on (PID2019-105544GB-I00, PID2022-140510NB-I00 and
RYC2019-027017-I), Xunta de Galicia (CIGUS Network of Research Centers,
Consolidaci\'on 2021 GRC GI-2033, ED431C-2021/22 and ED431F-2022/15),
Junta de Andaluc\'\i{}a (SOMM17/6104/UGR and P18-FR-4314), and the European
Union (Marie Sklodowska-Curie 101065027 and ERDF); USA -- Department of
Energy, Contracts No.~DE-AC02-07CH11359, No.~DE-FR02-04ER41300,
No.~DE-FG02-99ER41107 and No.~DE-SC0011689; National Science Foundation,
Grant No.~0450696, and NSF-2013199; The Grainger Foundation; Marie
Curie-IRSES/EPLANET; European Particle Physics Latin American Network;
and UNESCO.
\end{sloppypar}

% <!-----------------!> %

%Bibliography
%Citations can be done with cite: refs.~\cite{a,b,c}.

%\renewcommand*{\bibfont}{\small}
%\bibliographystyle{myunsrt}
%\bibliography{Bibliography}

% <!-----------------!> %

\renewcommand{\thetable}{A.\arabic{table}}
\renewcommand{\thefigure}{A.\arabic{figure}}
\setcounter{table}{0}
\setcounter{figure}{0}
%\clearpage
\appendix

\section{Lateral distribution function for photon events}
\label{sec:photonLDF}

The analytical model for the LDF used is based on the Nishimura-Kamata-Greisen (NKG) function~\citep{Kamata1958,Greisen1956}:

\begin{equation}
f_\text{LDF}(r) = S(r_{\rm ref}) \times \left( \frac{r}{r_{\rm ref}}  \times \frac{r+\unit[700]{m}}{r_{\rm ref}+\unit[700]{m}} \right)^{-\beta}
\label{eq:LDF}
\end{equation}

\noindent where $r_{\rm ref}=\unit[250]{m}$ in the case of SD-433 events. The slope parameter, $\beta$, is parametrized a priori as follows:

%where the reference distance, \mbox{$r_{\rm ref} = \unit[250]{m}$}, is chosen to minimize the uncertainties in the measured signal and is primarily determined by the \unit[433]{m} spacing between the stations~\citep{Newton2007}.
%Because air showers initiated by photon primaries develop almost purely through electromagnetic processes, the LDF steepness in photon-initiated events is expected to be greater than that in hadronic-initiated events, requiring fine-tuning. It is parameterized as follows:

\begin{equation}
\beta(S(250),\theta) = \beta_0(\theta) + \beta_1(\theta) \times \lg\left(\frac{S(250)}{f(\theta)}\right)
\label{eq:beta}
\end{equation}

\noindent using events with at least five triggered stations, none of which is saturated\footnote{A station is flagged as saturated when the digitized traces of the anode channel overflow.}, and between two and four stations located \unit[100]{m} to \unit[400]{m} from the shower axis. These criteria ensure that $\beta$ is obtained with at least two degrees of freedom in the event reconstruction and with topological coverage around \unit[250]{m}, providing a solid lever arm for the LDF fit.
%The criteria to reconstruct with a free β was adapted from a 1500-m array [50] to the distances of the SD-433.
%These events are used to determine the three factors in \cref{eq:beta}.

\begin{figure}[!tb]
\includegraphics[width=0.49\textwidth]{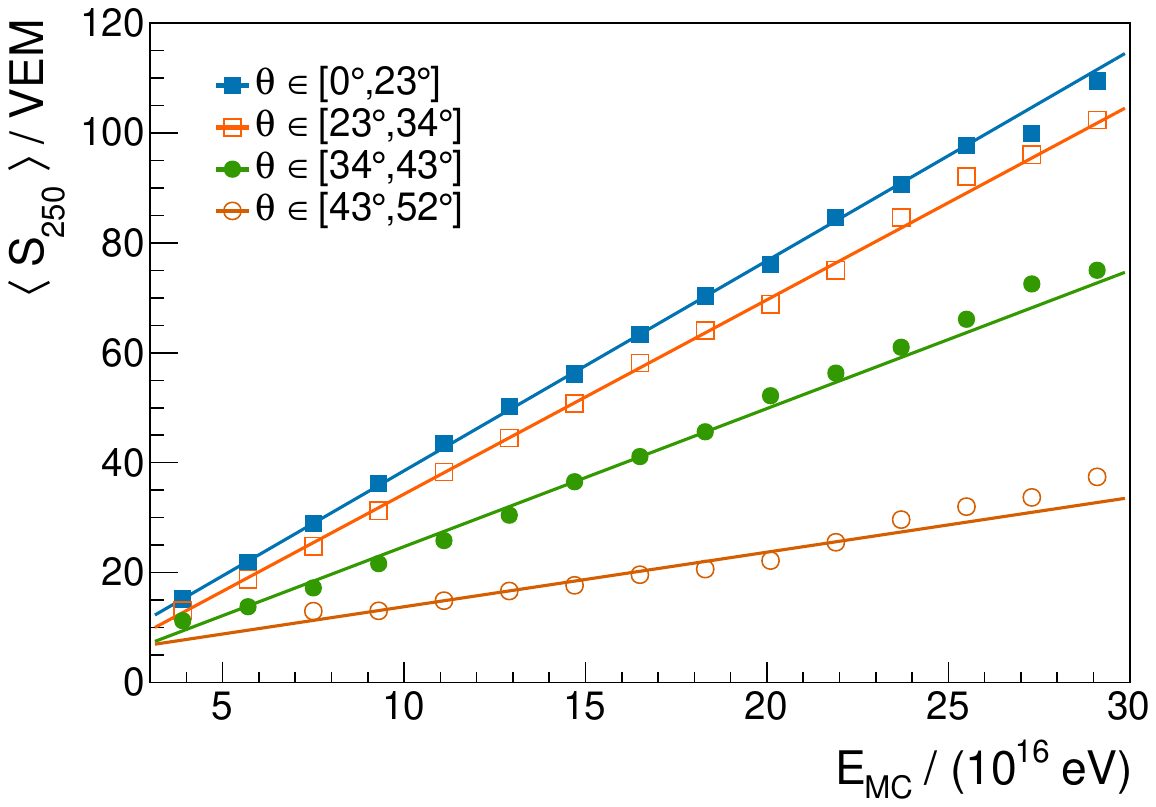} \includegraphics[width=0.49\textwidth]{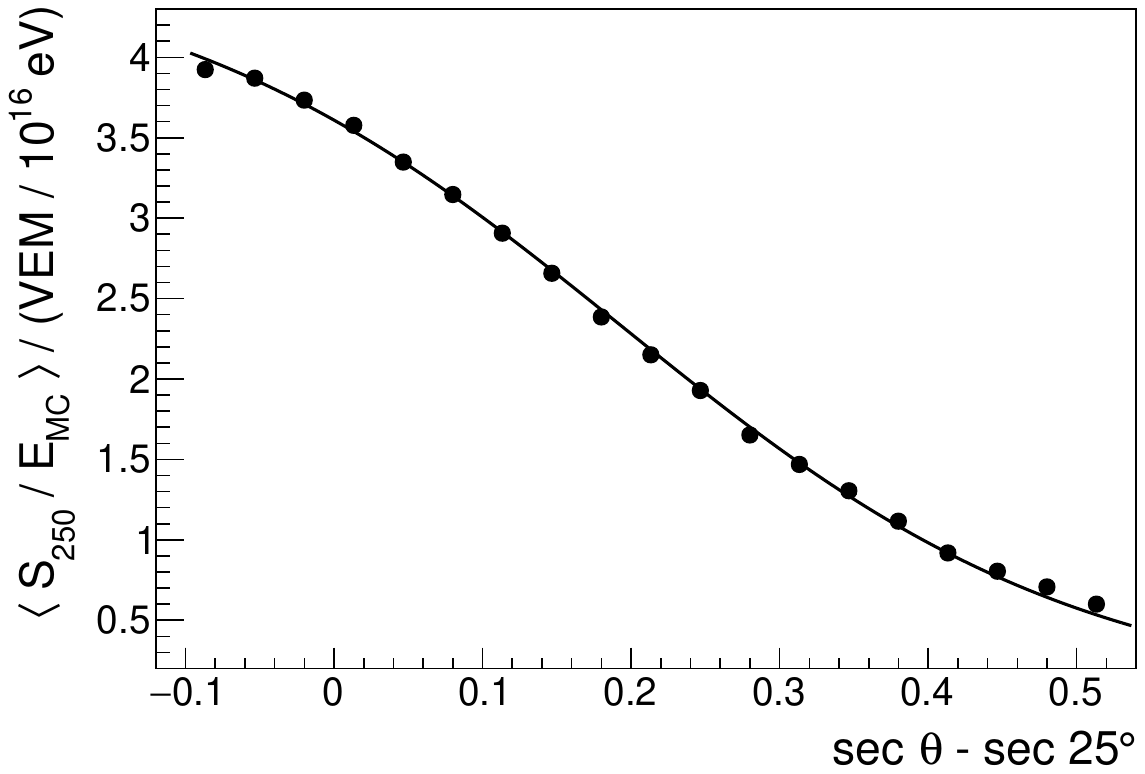}
\caption{Left: The shower size, $S(250)$, as a function of the simulated primary energy, $E_\text{MC}$, for photon events. The solid lines represent linear fits in zenith-angle bins. Right: The average ratio between the shower size and the simulated energy versus the zenith angle. The solid line corresponds to the fit $\tilde{f}(\theta)$ from \cref{eq:fattLdf} via a $\chi^2$ minimization.}
\label{fig:S250EMC}
\end{figure}

The function $f(\theta)$ compensates for the dependence of $S(250)$ on the zenith angle:

\begin{equation}
\label{eq:fattLdf}
f(\theta) = \tilde{f}(\theta)/ \tilde{f}(25^\circ)\text{, with }\tilde{f}(\theta) = \frac{f_0}{1+\text{exp}\left( \frac{x-b}{c} \right)} ,\,\text{ and }\, x = \sec\theta - \sec\unit[25]{^\circ}
\end{equation}
\noindent where the reference zenith angle, \unit[25]{$^\circ$}, is selected as the median of the $E_\text{MC}^{-2}$-weighted zenith distribution. The three free parameters of $\tilde{f}(\theta)$ are estimated by examining the approximately linear relationship between the $S_{250}$ and $E_\text{MC}$, as shown in \cref{fig:S250EMC}, left. Consequently, the ratio $S(250)/E_\text{MC}$ is mainly determined by the angular dependence, as depicted in \cref{fig:S250EMC}, right. The sigmoid function $\tilde{f}(\theta)$ is fitted to the average ratio, resulting in parameters \mbox{$f_0=\unit[(4.65\pm0.03)]{VEM/\unit[10^{16}]{eV}}$}, $b=(0.194\pm0.002)$ and $c=(0.156\pm0.002)$. This model captures the mean behavior within a $2\%$ margin.
%The correction factor accounting for the atmospheric attenuation, $f(\theta)$, is
%This correction is found by assuming that the energy and angular dependencies of $S(250)$ can be factorized.

\begin{figure}[!tb]
%\begin{wrapfigure}[14]{i}{0.5\textwidth}
\includegraphics[width=0.49\textwidth]{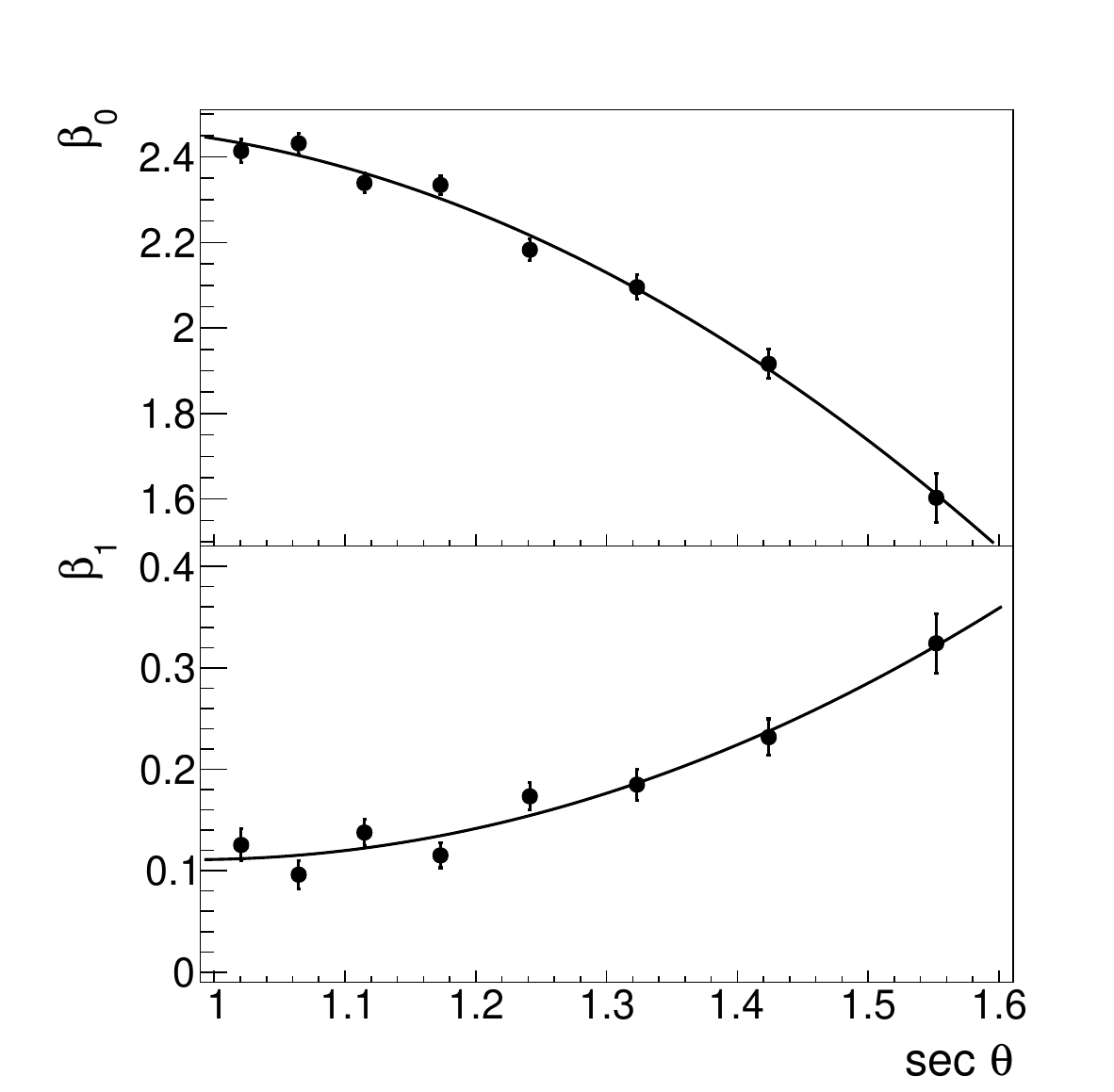} \includegraphics[width=0.49\textwidth]{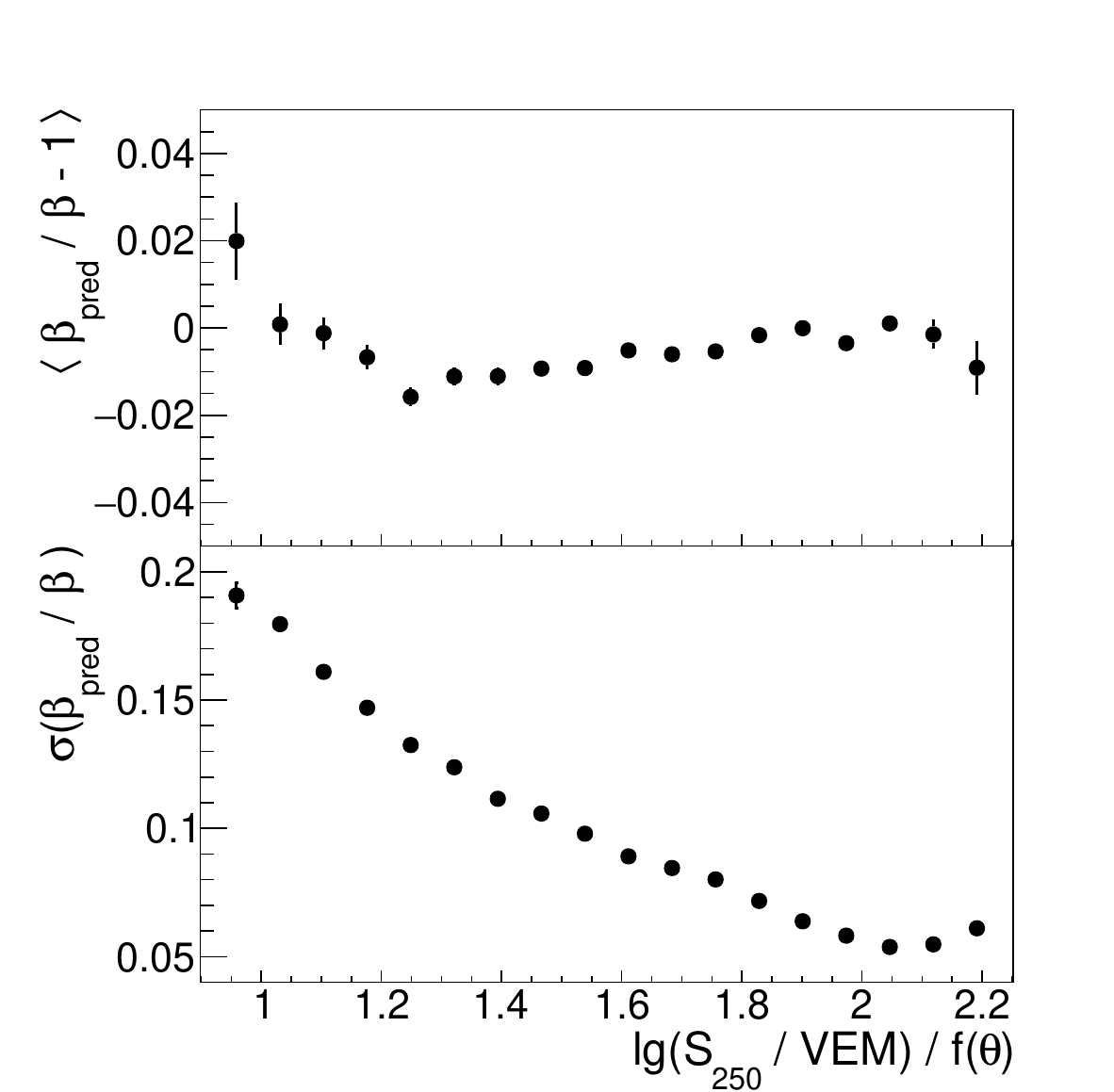}
\caption{Left: The evolution of $\beta_0$ and $\beta_1$ parameters of \cref{eq:beta} with $\sec\theta$. The second-order polynomials in \cref{eq:beta0beta1_ML} are superimposed. Right: The bias and resolution of the LDF slope predicted by \cref{eq:beta}, $\beta_\text{pred}$.}
\label{fig:betaParam}
%\end{wrapfigure}
\end{figure}

The linear coefficients, $\beta_0$ and $\beta_1$, depend quadratically on $\sec\theta$, as illustrated \mbox{in \cref{fig:betaParam}}, left. The corresponding six parameters are estimated using an unbinned maximum likelihood method, assuming a Gaussian probability density function for $\beta$. The resulting values are given in \cref{eq:beta0beta1_ML}. The performance of the parametrization is assessed by comparing the predicted slope from \cref{eq:beta}, $\beta_\text{pred}$, with the reconstructed slope for each photon-initiated event, $\beta$. The parametrization shows a negligible bias (within $\pm2\%$) and a resolution better than $19\%$ as illustrated in \cref{fig:betaParam}, right.

\begin{equation}
\label{eq:beta0beta1_ML}
\begin{aligned}
\beta_0 &= (0.756\pm0.011) + (4.05\pm0.01) \times \sec\theta - (2.33\pm0.01) \times \sec^2\theta\\
\beta_1 &= (0.704\pm0.006) - (1.37\pm0.01) \times \sec\theta + (0.758\pm0.004) \times \sec^2\theta
\end{aligned}
\end{equation}

\section{Parametrization of the background contamination and photon candidate cuts}
\label{sec:BC_and_CandCuts}

The average background contamination at $50\%$ signal efficiency and the photon candidate cut were parametrized in terms of the photon-equivalent energy scale. As shown by the dashed line in both panels of \cref{fig:BCCandCutCats}, the employed analytical models are:

\begin{equation}
\label{eq:BCE}
f_\text{BC}\left(E_{\gamma,\text{eq}}\right) = b_0 \times \text{e}^{-b_1 \times (\lg(E_{\gamma,\text{eq}}/\text{eV})-17)}
\end{equation}

\begin{equation}
\label{eq:photonCandCut}
f_\text{cut}\left(E_{\gamma,\text{eq}}\right) = - \left(c_0 + c_1 \times \left( \lg(E_{\gamma,\text{eq}}/\text{eV})-16 \right)^{-c_2} \right)
\end{equation}

The free parameters of both equations are given in \cref{tab:BCEPhotonCutEPars} for each event category.

\begin{table}[h]
\begin{center}
\begin{tabular}{c c c c c c}
\toprule
{\makecell{Event \\ category}} & {$b_0$} & {$b_1$} & {$c_0$} & {$c_1$} & {$c_2$} \\
\midrule
\midrule
{I} & {$(1.15\pm0.22)\times10^{-6}$} & {$6.20\pm0.50$} & {$0.262\pm0.035$} & {$1.04\pm0.28$} & {$32.5\pm8.5$} \\
{II} & {$(2.13\pm0.37)\times10^{-6}$} & {$6.76\pm0.51$} & {$0.306\pm0.020$} & {$1.03\pm0.12$} & {$30.8\pm4.2$} \\
{III} & {$(2.33\pm0.39)\times10^{-6}$} & {$6.19\pm0.50$} & {$0.411\pm0.012$} & {$1.28\pm0.22$} & {$39.3\pm4.6$} \\
{IV} & {$(1.05\pm0.15)\times10^{-5}$} & {$5.92\pm0.53$} & {$0.330\pm0.055$} & {$0.867\pm0.125$} & {$24.0\pm7.0$} \\
{V} & {$(1.87\pm0.35)\times10^{-6}$} & {$6.81\pm0.46$} & {$0.429\pm0.042$} & {$0.931\pm0.170$} & {$27.0\pm7.0$} \\
{VI} & {$(5.98\pm0.94)\times10^{-6}$} & {$5.77\pm0.43$} & {$0.470\pm0.018$} & {$1.10\pm0.17$} & {$34.2\pm4.7$} \\
\bottomrule
\end{tabular}
\end{center}
\caption{\label{tab:BCEPhotonCutEPars} The parameters of \cref{eq:BCE,eq:photonCandCut} modeling the background contamination at $50\%$ signal efficiency and the photon candidate cut for each event category.}
\end{table}

\end{document}